\documentclass[10pt]{article}  % onecolumn (standard format)
\usepackage[utf8]{inputenc}
\usepackage{amsmath}
\usepackage{amsfonts}
\usepackage{amssymb}
\usepackage{listings}
\usepackage{mathrsfs}
\usepackage{times}
\usepackage{float}
\usepackage{color}
\usepackage{setspace}
\usepackage{color,soul}

\usepackage{amsmath,amsfonts,amsthm,eucal}
\usepackage{enumerate}
\usepackage[letterpaper,margin=3cm]{geometry}
\usepackage{float}
\usepackage[title]{appendix}
\usepackage{color}
\usepackage{array}
\usepackage{comment}

\usepackage{multirow}
\usepackage{placeins}
\usepackage{bm}
\newcolumntype{M}[1]{>{\centering\arraybackslash}m{#1}}

\usepackage[
pdfstartview=XYZ,
bookmarks=true,
colorlinks=true,
linkcolor=blue,
urlcolor=blue,
citecolor=blue,
pdftex,
bookmarks=true,
linktocpage=true,   % makes the page number as hyperlink in table of content
hyperindex=true
]{hyperref}

\usepackage{lipsum}
\usepackage{lineno}
\usepackage{natbib}
\setcitestyle{authoryear}

\usepackage[pdftex]{graphicx}
\usepackage{epstopdf}
\usepackage{booktabs} 
\usepackage{lineno}
\usepackage{tikz,mathpazo}
\usetikzlibrary{shapes.geometric, arrows}
\usepackage{indentfirst}
\usepackage{graphicx}
\usepackage{subcaption}

\usepackage{caption}

\usepackage{algorithm}
\usepackage{algorithmicx}
\usepackage[noend]{algpseudocode}

\theoremstyle{remark}
\newtheorem{rmk}{Remark}

\usepackage{algorithm}
\usepackage[noend]{algpseudocode}

\title{ Denoising diffusion models for inverse design of inflatable structures with programmable deformations} 

\begin{document}

\author{
    Sara Karimi
    \and
    Nikolaos N. Vlassis \thanks{Corresponding author, Department of Mechanical and Aerospace Engineering, Rutgers University, Piscataway, NJ 08854. \textit{nick.vlassis@rutgers.edu}} 
}

\maketitle

\begin{abstract}
Programmable structures are systems whose undeformed geometries and material property distributions are deliberately designed to achieve prescribed deformed configurations under specific loading conditions. Inflatable structures are a prominent example, using internal pressurization to realize large, nonlinear deformations in applications ranging from soft robotics and deployable aerospace systems to biomedical devices and adaptive architecture. 
We present a generative design framework based on denoising diffusion probabilistic models (DDPMs) for the inverse design of elastic structures undergoing large, nonlinear deformations under pressure-driven actuation. 
The method formulates the inverse design as a conditional generation task, using geometric descriptors of target deformed states as inputs and outputting image-based representations of the undeformed configuration.
Representing these configurations as simple images is achieved by establishing a pre- and postprocessing pipeline that involves a fixed image processing, simulation setup, and descriptor extraction methods.
Numerical experiments with scalar and higher-dimensional descriptors show that the framework can quickly produce diverse undeformed configurations that achieve the desired deformations when inflated, enabling parallel exploration of viable design candidates while accommodating complex constraints. 
\end{abstract}

\section{Introduction}
\label{sec:intro}

Programming elastic structures to achieve large, target deformations is central to applications requiring controlled shape change \citep{oliver2016morphing, zhou2016reversible}, adaptive behavior \citep{fan2016review, walther2020responsive}, or mechanical functionality \citep{mccracken2020materials, Fang2022-xf}.
Here, "programming" refers to the deliberate design of undeformed geometries and material property distributions that produce prescribed deformed configurations under specific loading conditions. 
Mechanical metamaterials exemplify this principle: their unit cell geometries are tailored to produce pattern transformations that yield negative Poisson’s ratio \citep{ren2018auxetic}, tunable stiffness \citep{Bertoldi2017-lq},  or controlled energy absorption \citep{fu2019programmable}. 
Deployable structures in aerospace applications \citep{Zirbel2013-zg, santo2014shape, zhang2021deployable} -- such as foldable antennas, booms, or membrane-based enclosures -- are similarly designed to transition from compact stowed states to precise operational geometries through elastic deformation. 
In stretchable electronics and wearable sensors \citep{Rogers2010-mf, homayounfar2020wearable}, substrates must undergo significant deformation while maintaining electrical and mechanical integrity \citep{Kim2008-pt}. 
Across these systems, the key challenge lies in mapping undeformed configurations to desired deformed states in a way that accounts for geometric and material nonlinearities, complex loading, and boundary constraints \citep{zheng2023deep,jiao2023mechanical}.

A prominent subset of such systems includes pneumatically actuated or inflatable structures \citep{ali2009review, fenci2017deployable, xavier2022soft}, which use internal pressurization to drive deformation and mechanical response.
In soft robotics, elastomeric actuators with embedded air chambers bend \citep{Shepherd2011-jj}, elongate \citep{10.1016/j.robot.2014.08.014}, or twist \citep{Connolly2015-qy} in response to applied pressure, enabling locomotion, grasping, or shape adaptation \citep{Rus2015-ue}. 
The structure of these actuators guides their deformation \citep{Gorissen2020-ii}. 
For example, they often incorporate anisotropic reinforcements -- e.g., helically wound fibers -- to constrain expansion and produce targeted motions \citep{Bishop-Moser2015-lu}. 
In biomedical applications,  balloon-expandable stents \citep{dumoulin2000mechanical}, inflatable catheters \citep{choi2014novel}, and occlusion devices \citep{gianturco1975mechanical} deploy from compact states to functional geometries within the body, requiring precise shape change under physiological constraints \citep{Timmins2008-jy}. 
Similarly, architectural systems, such as deployable shelters \citep{Moy2022-wz} and ETFE cushions \citep{Hu2017-sw},  rely on controlled inflation to form tension-stabilized surfaces or load-bearing arches. 
The common thread in these applications is the targeted use of pressure-driven actuation to realize specific deformed geometries to perform given tasks, often under non-uniform constraints and with strong coupling between material properties, geometry, and loading paths \citep{skouras2014designing,gorissen2017elastic}.

Conventional approaches to inverse design of deformable structures -- such as topology optimization \citep{Bendsoe1988-qf} or heuristic iterative schemes \citep{Xie1993-lt} -- typically require repeated calls to finite element method (FEM) simulations  \citep{Rozvany2001-oy, Liu2014-bd}, making them computationally prohibitive for large-scale design spaces \citep{mukherjee2021accelerating}. These methods often rely on simplified material models \citep{Andreassen2011-ka} or gradient-based updates that can struggle with non-smooth or highly nonlinear responses, particularly under large deformations \citep{Bruns2001-ps}. Moreover, most existing frameworks assume simpler boundary conditions and domains shapes or have to adopt advanced methods to handle complex configurations \citep{rozvany2012structural, Wang2003-en, yoon2011topology, guo2013robust}, limiting their applicability in systems with complex actuation, such as pressure-driven inflatable structures.
The final deformed configurations result from complex interactions among initial geometry, boundary constraints, and material properties, often under non-uniform pressure loads and contact conditions \citep{caasenbrood2020computational, panetta2021computational}, which can violate assumptions commonly made in conventional design pipelines.

To address these limitations, data-driven inverse design frameworks have emerged as alternatives to classical computational methods \citep{regenwetter2022deep, koul2024review, cerniauskas2024machine}. Many of these leverage latent space generative models, such as variational autoencoders (VAEs) \citep{kingma2013auto} and generative adversarial networks (GANs) \citep{goodfellow2014generativeadversarialnetworks} among others, to learn mappings between design parameters and structural responses from simulated or experimental datasets. 
VAEs offer stable training dynamics and structured latent spaces, making them appealing for inverse design tasks \citep{kim2021exploration, yonekura2021data}. However, they often require strong regularization to enforce smoothness and disentanglement in the latent space \citep{mathieu2019disentangling}, which can reduce the fidelity of reconstructed or generated samples and may struggle to capture sharp, high-frequency features \citep{asperti2020balancing} such as localized  material discontinuities.
GAN-based models, on the contrary, can generate sharper and more detailed outputs, making them attractive for applications where fine geometric fidelity is critical \citep{yang2018microstructural,mao2020designing}.
However, they are prone to training instabilities, including mode collapse and sensitivity to hyperparameters arising from the need to maintain a delicate balance between generator and discriminator \citep{saxena2021generative}.
Recent studies attempt to address these limitations by incorporating physics-informed loss functions and hybrid discriminator designs \citep{kazemi2022multiphysics, nguyen2022synthesizing}.

In parallel, denoising diffusion probabilistic models (DDPMs) \citep{DBLP:journals/corr/abs-2006-11239} have been replacing state-of-the-art methods in synthesis tasks and, by extension, have gained traction in the inverse design of structures \citep{vlassis2023denoising, Bastek2023-il, kartashov2025large} and the development of digital twins \citep{Kadry2024-ed,vlassis2024synthesizing}.  
DDPMs offer several advantages over earlier generative models: they exhibit stable training dynamics, produce high-quality and diverse samples, and allow for explicit control over the generation process through conditioning mechanisms \citep{croitoru2023diffusion, yang2023diffusion}.  
DDPMs offer a more stable and diverse, while often slower, generation alternative to GANs \citep{dhariwal2021diffusion, bayat2023study} by progressively refine noisy samples into data-like outputs through a well-defined iterative denoising process. This makes them particularly well-suited for complex design spaces and tasks requiring fine-grained control.

In this work, we develop a generative design framework based on DDPMs to tackle the inverse design of elastic structures undergoing large, nonlinear deformations under pressure-driven actuation. 
The framework generates structural designs that deform into prescribed geometries when inflated under fixed boundary conditions.
Instead of relying on explicit representations of meshes or simulation graphs, the method operates directly on simpler image-based inputs and outputs by setting up fixed pre- and post-processing pipelines. 
We formulate the design problem as a conditional generation task, where geometric descriptors computed from simulated deformations (e.g., height, width, convex hull area, or medial axis of the deformed configuration) serve as the conditioning inputs. 
To validate the framework, we perform extensive numerical experiments across a range of scalar and higher-dimensional descriptors and evaluate the ability of the generated designs to match target deformation metrics using high-fidelity finite element simulations. 

This framework enables the rapid generation of large numbers of undeformed configurations that achieve prescribed size- and shape-based descriptors when inflated, supporting functional targets in inflatable structures. 
It also aims to streamline the treatment of complex boundary conditions inherent in such problems and simplify the representation of candidate structures.
By producing many viable designs with equivalent performance, the approach aims to facilitate massive parallel prototyping and offer flexibility in selecting configurations that best meet additional manufacturing or operational constraints.

This paper is organized as follows. Section \ref{sec:theoretical_background} introduces the diffusion-based inverse design framework, including the forward and reverse processes, conditioning strategy, and training objective. Section \ref{sec:database} describes the construction of the simulation dataset, covering preprocessing of undeformed structures, boundary condition specification, finite element analysis, descriptor extraction, and boundary condition–preserving data augmentation. 
Section \ref{sec:numerical_experiments} outlines the training configuration and presents five numerical experiments evaluating the framework’s ability to generate undeformed structures that deform to match target geometric descriptors of varying complexity. Section \ref{sec:conclusion} summarizes the main findings and discusses potential extensions to broader classes of inverse design problems.

\section{Theoretical Background}
\label{sec:theoretical_background}

This section presents the theoretical foundations of the proposed generative framework based on denoising diffusion probabilistic models.
We first describe the forward noising process, which defines a fixed Markov chain that progressively perturbs structural basis images using Gaussian noise. 
We then formulate the conditional reverse process, which is learned by a U-Net model to generate structural configurations conditioned on target deformation descriptors, along with the context embedding strategy that integrates both geometric and temporal information. 
The hybrid training objective combining a noise-prediction loss and a variational lower bound is also described.
Finally, we outline the complete DDPM pipeline for programmable deformation, covering both the training and sampling workflows.

\subsection{Forward Process}
\label{sec:fp}

The generative diffusion framework begins with a forward process that progressively perturbs clean data samples of structural bases using small amounts of Gaussian noise. 
A structural basis $\mathbf{x}$ is defined as a 2D image representation of the undeformed configuration discussed in more detail in Section~\ref{sec:database}. 
Representing the undeformed configuration as an image allows to closely follow an established formulation of the DDPMs that is common in the literature. Specifically, in this work we adapt the implementations of DDPMs introduced in \cite{DBLP:journals/corr/abs-2006-11239} and \cite{nichol2021improved}.
While we present the necessary formulations used in this work, the reader is directed to these works for a more detailed derivation of the DDPM framework. 

Given a dataset with initial samples drawn from a distribution \( q(\mathbf{x}_0) \), each sample \( \mathbf{x}_0 \) undergoes a gradual noising process across \( T \) discrete diffusion timesteps to produce a sequence \( \mathbf{x}_1, \dots, \mathbf{x}_T \). 
At each step \( t \), Gaussian noise with a small variance selected from a scalar variance schedule \( \beta_t \in (0, 1) \) is added.
This will lead to a complete loss of the structural basis signal in \( \mathbf{x}_T \) when \( T \) theoretically approaches infinity -- in practice, a large number of steps is usually selected. 
In this work, for example, we select $T= 1000$.

This forward noising process $q$ can be decribed as a Markov process -- each diffusion depends only on the previous step given the noising schedule. 
This Markov process is defined by a chain of conditional distributions:
\begin{equation}
q(\mathbf{x}_1, \dots, \mathbf{x}_T \mid \mathbf{x}_0) = \prod_{t=1}^T q(\mathbf{x}_t \mid \mathbf{x}_{t-1}),
\end{equation}
where each transition is governed by:
\begin{equation}
q(\mathbf{x}_t \mid \mathbf{x}_{t-1}) = \mathcal{N}(\mathbf{x}_t; \sqrt{1 - \beta_t} \, \mathbf{x}_{t-1}, \beta_t \mathbf{I}).
\end{equation}

By defining \( \alpha_t = 1 - \beta_t \) and the cumulative product \( \bar{\alpha}_t = \prod_{s=1}^t \alpha_s \), the marginal distribution at timestep \( t \) can be expressed directly in terms of the original sample:
\begin{equation}
q(\mathbf{x}_t \mid \mathbf{x}_0) = \mathcal{N}(\mathbf{x}_t; \sqrt{\bar{\alpha}_t} \, \mathbf{x}_0, (1 - \bar{\alpha}_t) \mathbf{I}).
\end{equation}
The following reparameterization trick gives a closed-form expression for generating noisy samples at any step:
\begin{equation}
\mathbf{x}_t = \sqrt{\bar{\alpha}_t} \, \mathbf{x}_0 + \sqrt{1 - \bar{\alpha}_t} \, \boldsymbol{\epsilon}, \quad \boldsymbol{\epsilon} \sim \mathcal{N}(0, \mathbf{I}),
\end{equation}
where $\epsilon$ is a random noise variable.

In addition, the posterior distribution of \( \mathbf{x}_{t-1} \) given \( \mathbf{x}_t \) and \( \mathbf{x}_0 \) is also Gaussian:
\begin{equation}
q(\mathbf{x}_{t-1} \mid \mathbf{x}_t, \mathbf{x}_0) = \mathcal{N}(\mathbf{x}_{t-1}; \tilde{\boldsymbol{\mu}}_t(\mathbf{x}_t, \mathbf{x}_0), \tilde{\beta}_t \mathbf{I}),
\end{equation}
where the mean and adjusted variance are given by:
\begin{align}
\tilde{\boldsymbol{\mu}}_t(\mathbf{x}_t, \mathbf{x}_0) &= \frac{\sqrt{\bar{\alpha}_{t-1}} \, \beta_t}{1 - \bar{\alpha}_t} \, \mathbf{x}_0 + \frac{\sqrt{\alpha_t} (1 - \bar{\alpha}_{t-1})}{1 - \bar{\alpha}_t} \, \mathbf{x}_t, \\
\tilde{\beta}_t &= \frac{1 - \bar{\alpha}_{t-1}}{1 - \bar{\alpha}_t} \, \beta_t.
\end{align}

This forward (noising) process is a fixed Markov chain determined by the chosen variance schedule. It serves as the basis for the reverse process, which is learned by the model to approximate the denoising transitions, as detailed below.

\subsection{Conditional reverse process for structure basis generation}
\label{sec:crf}

The conditional reverse diffusion process enables the generation of undeformed structures that are guided by specified geometric descriptors of the target deformed state. 
In this setting, the model is trained to reverse the forward noising process, producing structural bases that, under fixed boundary conditions, deform to match prescribed descriptors.

Let $\mathbf{y}$ represent the conditioning descriptor associated with a target deformation. This vector $\mathbf{y}$ is first passed through an embedding network $\boldsymbol{\zeta}_d(\mathbf{y})$ that maps it into a latent  space representation. 
This  deformation context embedded feature vector $\boldsymbol{\zeta}_d$ is added to a time step embedded feature vector $\boldsymbol{\zeta}_t$ -- that is calculated through sinusoidal positional embedding and holds information about the current denoising step $t$. 
The total context embedded feature vector is then $\boldsymbol{\zeta}(\mathbf{y},t) = \boldsymbol{\zeta}_t (t) + \boldsymbol{\zeta}_d(\mathbf{y})$.

It can be shown that for very small variance $\beta_t$ or very large number of diffusion steps $T$, the reverse of the diffusion process, a denoising process $p$ is a Markov process that has the same form as the forward process $q$ \citep{feller2015}. 
In order to achieve the targeted generation, the reverse process has to be conditioned on the target deformed shape descriptors. 
Thus, we can model the denoising process as a product of conditional Gaussians:
\begin{equation}
p\left(\mathbf{x}_{0: T},\mathbf{y}\right)=p\left(\mathbf{x}_T\right) \prod_{t=1}^T p\left(\mathbf{x}_{t-1} \mid \mathbf{x}_t,\boldsymbol{\zeta}(\mathbf{y},t)\right),
\end{equation}
to generate a coherent sample start from $p\left(\mathbf{x}_T\right)$, a random noise distribution sampled as $\mathcal{N}\left(x_T ; 0, I\right)$, given target deformation descriptor $\mathbf{y}$. 
Calculating such a reverse process $p$ would be computationally prohibitive, so it is approximated by a neural network $p_\theta$.

The reverse process $p_\theta$ can also be modeled as a Gaussian distribution such that:
\begin{equation}
p_\theta(\mathbf{x}_{t-1} | \mathbf{x}_t, \boldsymbol{\zeta}(\mathbf{y},t)) = \mathcal{N}\left(\mu_\theta(\mathbf{x}_t, \boldsymbol{\zeta}(\mathbf{y},t)), \Sigma_\theta(\mathbf{x}_t, t)\right),
\end{equation}
where the mean can be learned by the model and is given by
\begin{equation}
  \mu_\theta(\mathbf{x}_t, \boldsymbol{\zeta}(\mathbf{y},t), t) = \frac{1}{\sqrt{\alpha_t}} \left( \mathbf{x}_t - \frac{1 - \alpha_t}{\sqrt{1 - \bar{\alpha}_t}} \boldsymbol{\epsilon}_\theta(\mathbf{x}_t, \boldsymbol{\zeta}(\mathbf{y},t), t) \right),
  \label{eq:mu}
\end{equation}
and the variance 
is set $\Sigma_\theta\left(\mathbf{x}_t, t\right)=\sigma_t^2 \mathbf{I}$ with $\sigma_t^2=\beta_t$ as untrained time dependent constants. In practice, the network is tasked to predict noise $\boldsymbol{\epsilon}_\theta$ to remove -- implicitly approximating the mean $\mu_\theta$ as shown in Eq~\eqref{eq:mu}. In our framework, we use a U-Net architecture \citep{ronneberger2015unetconvolutionalnetworksbiomedical} to perform the denoising by predicting the noise component $\boldsymbol{\epsilon}_\theta(\mathbf{x}_t, \boldsymbol{\zeta}(\mathbf{y},t))$ at each denoising step $t$.

The model is trained using a hybrid learning objective that combines the simplified denoising loss introduced in  \citet{DBLP:journals/corr/abs-2006-11239} with a weighted variational lower bound (VLB) term, following the formulation of \citet{nichol2021improved}:
\begin{equation}
L_{\text{hybrid}} = L_{\text{simple}} + \lambda L_{\text{vlb}},
\label{eq:hybrid_loss}
\end{equation}
where $\lambda$ controls the contribution of the VLB term. The first term is the standard epsilon prediction loss:
\begin{equation}
L_{\text{simple}} := \mathbb{E}_{t, \mathbf{x}_0, \boldsymbol{\epsilon}}\left[\left\|\boldsymbol{\epsilon} - \boldsymbol{\epsilon}_\theta\left(\sqrt{\bar{\alpha}_t} \mathbf{x}_0 + \sqrt{1 - \bar{\alpha}_t} \boldsymbol{\epsilon}, t\right)\right\|^2\right],
\end{equation}
and the second term corresponds to a discretized variational lower bound:
\begin{equation}
L_{\text{vlb}} := \sum_{t=0}^T L_t,
\end{equation}
with
\begin{equation}
L_t := \begin{cases}
   - \log p_\theta(\mathbf{x}_0|\mathbf{x}_1), & \text{if } t = 0, \\
   D_{\text{KL}}\left(q(\mathbf{x}_{t-1}|\mathbf{x}_t, \mathbf{x}_0) \,\|\, p_\theta(\mathbf{x}_{t-1}|\mathbf{x}_t)\right), & \text{if } 0 < t < T, \\
   D_{\text{KL}}\left(q(\mathbf{x}_T|\mathbf{x}_0) \,\|\, p(\mathbf{x}_T)\right), & \text{if } t = T.
\end{cases}
\end{equation}
Each term in the above optimizes the Kullback–Leibler (KL) divergence between Gaussian distributions at successive denoising steps:
\begin{equation}
D_{\text{KL}}\left(p_\theta \| q\right) = \int_x p_\theta(x) \log \frac{p_\theta(x)}{q(x)}.
\end{equation}

It is noted that all parameters of the U-Net architecture that performs the denoising and the parameters of the descriptor embedding module $\boldsymbol{\zeta}_d$ that encodes the target geometric features are optimized jointly through backpropagation using the loss in Eq.~\eqref{eq:hybrid_loss}.

\subsection{DDPM framework for programmable deformation}
\label{sec:ddpm-programmable}

\begin{figure}[H]
    \centering
    \includegraphics[width=1\linewidth]{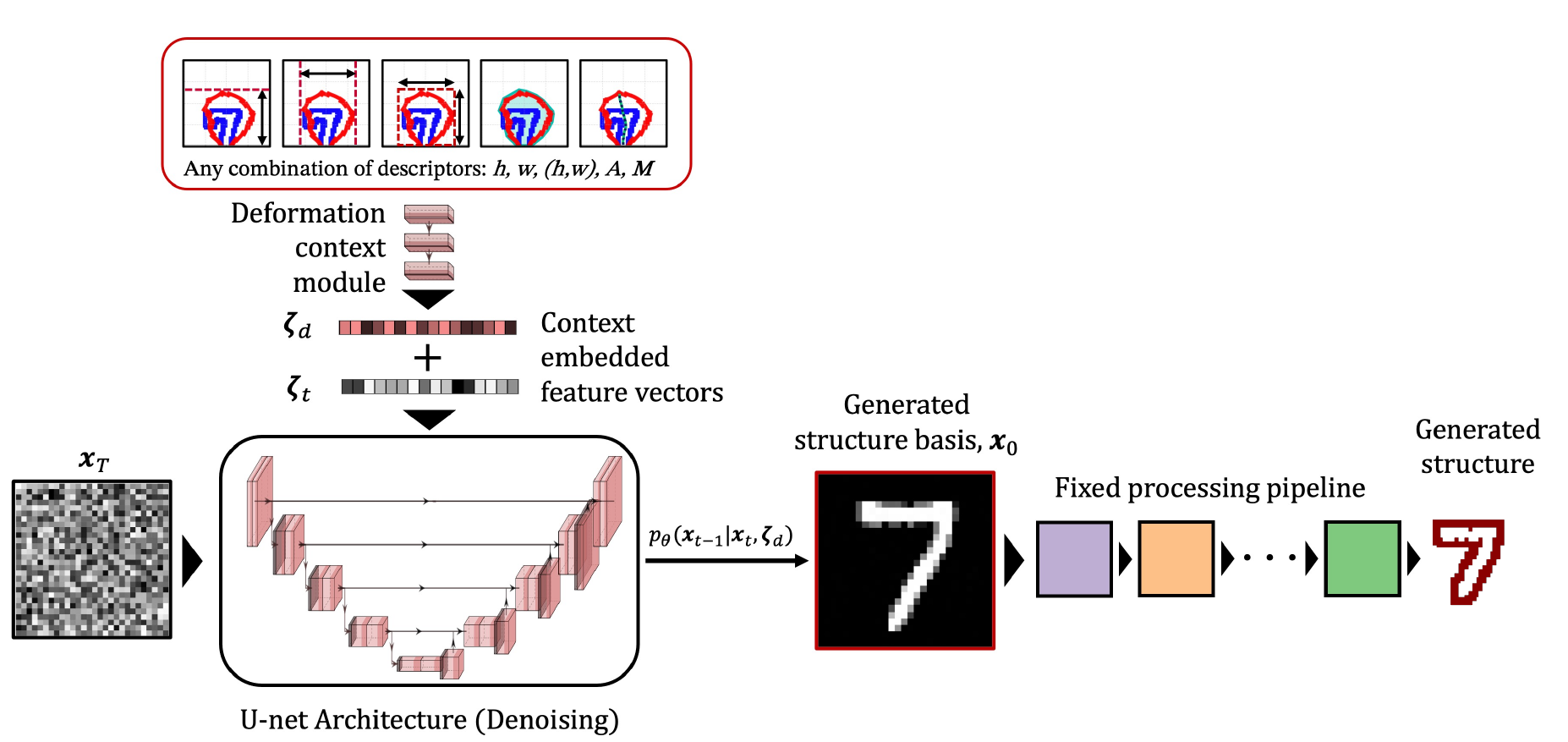}
    \caption{Schematic of the generative framework for the design of inflatable structures with target deformations.}
    \label{fig:schematic}
\end{figure}

Fig.~\ref{fig:schematic} illustrates the overall workflow of the proposed diffusion-based generative framework for programmable deformation. The schematic shows both training and sampling phases. 
During training, a structural basis image $\mathbf{x}_0$ is perturbed using the forward diffusion process to obtain a noisy sample $\mathbf{x}_t$ at a randomly selected timestep $t$. 
Simultaneously, a geometric descriptor $\mathbf{y}$ -- representing the target deformation -- is passed through an embedding network to produce a context vector $\boldsymbol{\zeta}_d(\mathbf{y})$. 
This is combined with a sinusoidal embedding of the current timestep $\boldsymbol{\zeta}_t(t)$ to obtain the full conditioning input $\boldsymbol{\zeta}(\mathbf{y}, t)$, which is provided to the U-Net denoising model. 
The model is trained to predict the noise $\boldsymbol{\epsilon}$ added during the forward process, enabling supervised learning via the hybrid loss described above.

During sampling, the learned reverse process is initialized with a sample drawn from a standard Gaussian distribution, $\mathbf{x}_T \sim \mathcal{N}(0, \mathbf{I})$. 
At each denoising step $t$, the descriptor $\mathbf{y}$ is embedded and combined with the sinusoidal time embedding to generate $\boldsymbol{\zeta}(\mathbf{y}, t)$, which conditions the U-Net model. 
The model predicts the noise to remove, producing a sequence of denoised estimates $\mathbf{x}_{T}, \mathbf{x}_{T-1}, \dots, \mathbf{x}_0$. 
The final output $\mathbf{x}_0$ is a coherent structural basis sample consistent with the target deformation descriptor. 
This generated image is then passed through a fixed preprocessing pipeline to be converted into a finite element model, which can be simulated under fixed boundary conditions to verify the target deformation as described in the following section.

\section{Database}
\label{sec:database}

In this section, we describe the construction of a standardized dataset used to train and evaluate the generative diffusion framework for inverse structural design under deformation constraints. The dataset consists of a collection of structural bases -- undeformed configurations encoded as grayscale images -- and their corresponding deformation descriptors, obtained through simulation under fixed boundary and loading conditions. 
All samples are processed using a consistent pipeline involving image preprocessing, meshing, simulation, and postprocessing. This standardization ensures compatibility with the simulation environment and allows the model to implicitly learn the structure–response relationships. We apply data augmentation that expand the design space while explicitly maintaining consistency with the applied boundary conditions.

\subsubsection{Fixed preprocessing pipeline and simulation}
The structural basis represents the undeformed configuration of a candidate design and is encoded as a grayscale image. In this study, structural bases are derived from a subset of the MNIST database -- a benchmark dataset of 
$28\times 28$ pixel images of handwritten digits commonly used in computer vision and generative modeling. We focus exclusively on images labeled as the digit “7” because they exhibit simple, continuous geometries that resemble the shape of soft robotic limbs, actuators, or inflatable components. These images serve as convenient design representations: they are small in size, topologically simple, and defined on a fixed grid, which facilitates meshing and downstream processing. Treating them as structural inputs allows for direct integration with diffusion-based generative models that operate in the image domain. All subsequent simulation and generation tasks are built around this representation of the undeformed structure.

\begin{figure}[H]
    \centering
    \includegraphics[width=0.97\linewidth]{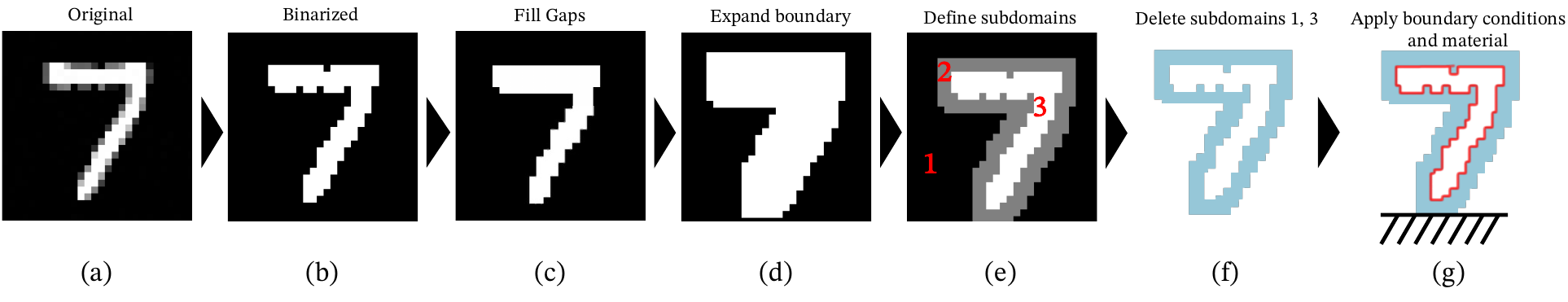}
    % \captionsetup{justification=raggedright,singlelinecheck=false}
    \caption{Fixed preprocessing pipeline of the structural bases images that represent the undeformed configuration.}
    \label{fig:placeholder}
\end{figure}

Each structural basis image is processed through a fixed image preprocessing pipeline to produce a meshable, physically interpretable geometry ready for FEM simulations. 
The steps of this pipeline are illustrated in Fig.~\ref{fig:placeholder}. 
The input is a grayscale image (Fig.~\ref{fig:placeholder}a), which is binarized to isolate the main connected component (Fig.~\ref{fig:placeholder}b). 
Binarization is used because pixel intensity is not treated as a continuous variable -- different fixed values will be assigned to distinct material domains in the simulation. 
Next, we fill interior gaps in the binarized region, if they are present, to remove pixel-scale discontinuities and ensure connectivity across the grid (Fig.~\ref{fig:placeholder}c). 
The undeformed configuration is then constructed by expanding the boundary of the digit by two pixels in all directions using a morphological maximum filter. 
The expanded geometry is assigned a pixel value of 0.5, while the original digit region is overlaid on top and set to 1 (Fig.~\ref{fig:placeholder}d). 
The resulting composite image (Fig.~\ref{fig:placeholder}e) defines the simulation subdomains, separating solid material regions from internal voids. 
Finally, any non-zero pixel value image content is shifted downward until it reaches the bottom edge of the frame (Fig.~\ref{fig:placeholder}f). 
This ensures a consistent reference location for applying boundary conditions and assigning material properties %in the subsequent simulation stage
(Fig.~\ref{fig:placeholder}g).

The final processed image (Fig.~\ref{fig:placeholder}g) is meshed and simulated through the Multiphysics Object-Oriented Simulation Environment (MOOSE) \citep{giudicelli2024moose} to compute the deformation behavior of the structure under internal pressurization. 
The grayscale image is interpreted as a labeled domain: pixels with value 0 correspond to the external void (background), pixels with value 1 define the internal void region, and pixels with value 0.5 represent the solid material. 
During mesh construction, the solid domain -- corresponding to pixels with value 0.5 -- is retained. It is assigned neo-Hookean material properties with Lamé parameters $\lambda = 16.44$ and $\mu = 0.1644$, where  $\lambda$ is the first Lamé parameter and $\mu$ is the shear modulus.
Both void regions are deleted from the mesh.
This results in a closed structural boundary enclosing the internal void. 

The simulation is executed in a transient setting using a total Lagrangian formulation with large deformations. 
A time-dependent internal pressure is applied along the interface between the solid material and the removed internal void region, defined as $p(t) = 0.01\times t$. 
This produces a smooth inflation response over time. 
The simulation proceeds from $t = 0$ to $t = 1$ with a time step size of $\Delta t = 0.1$. 
The bottom edge of the solid structure is fixed in both the horizontal and vertical directions to eliminate rigid-body motion. 
At each time step, the displacement field, components of the first Piola–Kirchhoff stress tensor, and components of the deformation gradient are computed and recorded. 
In this work, we focus exclusively on the displacement field at the final simulation time step, from which the deformed shape is extracted and used to compute geometric descriptors. The remaining quantities, such as stress and strain measures, as well as intermediate deformation states, are retained for potential use in future work involving multi-objective inverse design.
As such, we ran 6265 inflation simulations, equal to the number of "7" digit images in the MNIST dataset, to generate a diverse set of structure–response pairs for training and evaluation of the generative framework. This dataset will be augmented as discussed in the following sections.

\subsection{Postprocessing and descriptor extraction}

To simplify the representation of the deformed shapes used as targets for generation, we employ a set of geometric size and shape descriptors. 
These descriptors provide a compact summary of each structure’s response, making them suitable for conditioning generative models. 
The selected descriptors are representative and computationally inexpensive to extract, but they are not unique; alternative descriptors could be used depending on the desired resolution, interpretability, or application context. 
Our choices prioritize generality, ease of computation, and relevance to shape characterization across diverse geometries.

\captionsetup[subfigure]{labelformat=simple, labelsep=none, skip=0pt}
\renewcommand\thesubfigure{(\alph{subfigure})}

\begin{figure}[h!]
    \centering
    \begin{subfigure}[b]{0.19\textwidth}
        \includegraphics[height=3.6cm]{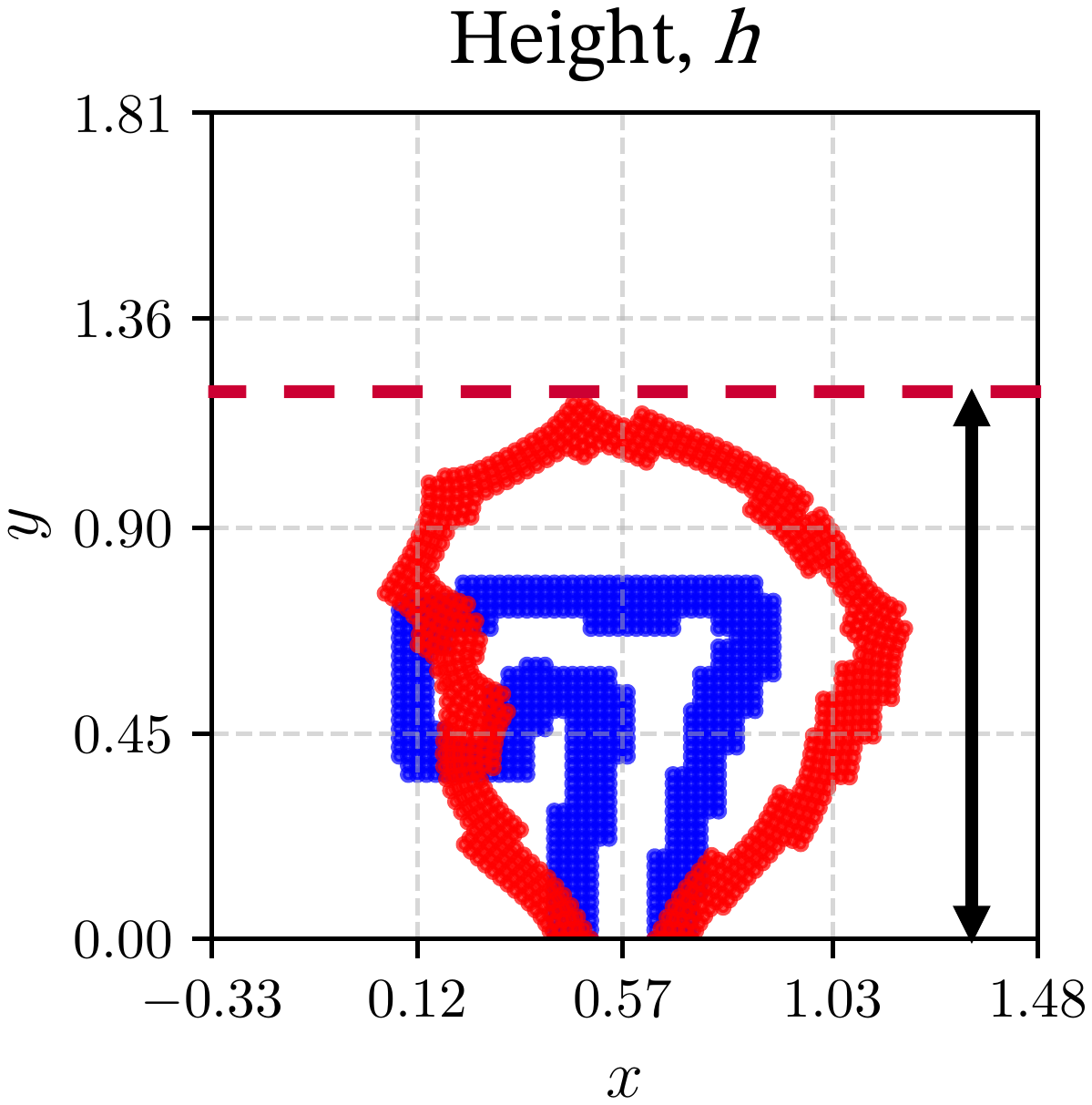}
        \caption{}
        \label{fig:desc_heights}
    \end{subfigure}
    \hfill
    \begin{subfigure}[b]{0.19\textwidth}
        \includegraphics[height=3.6cm]{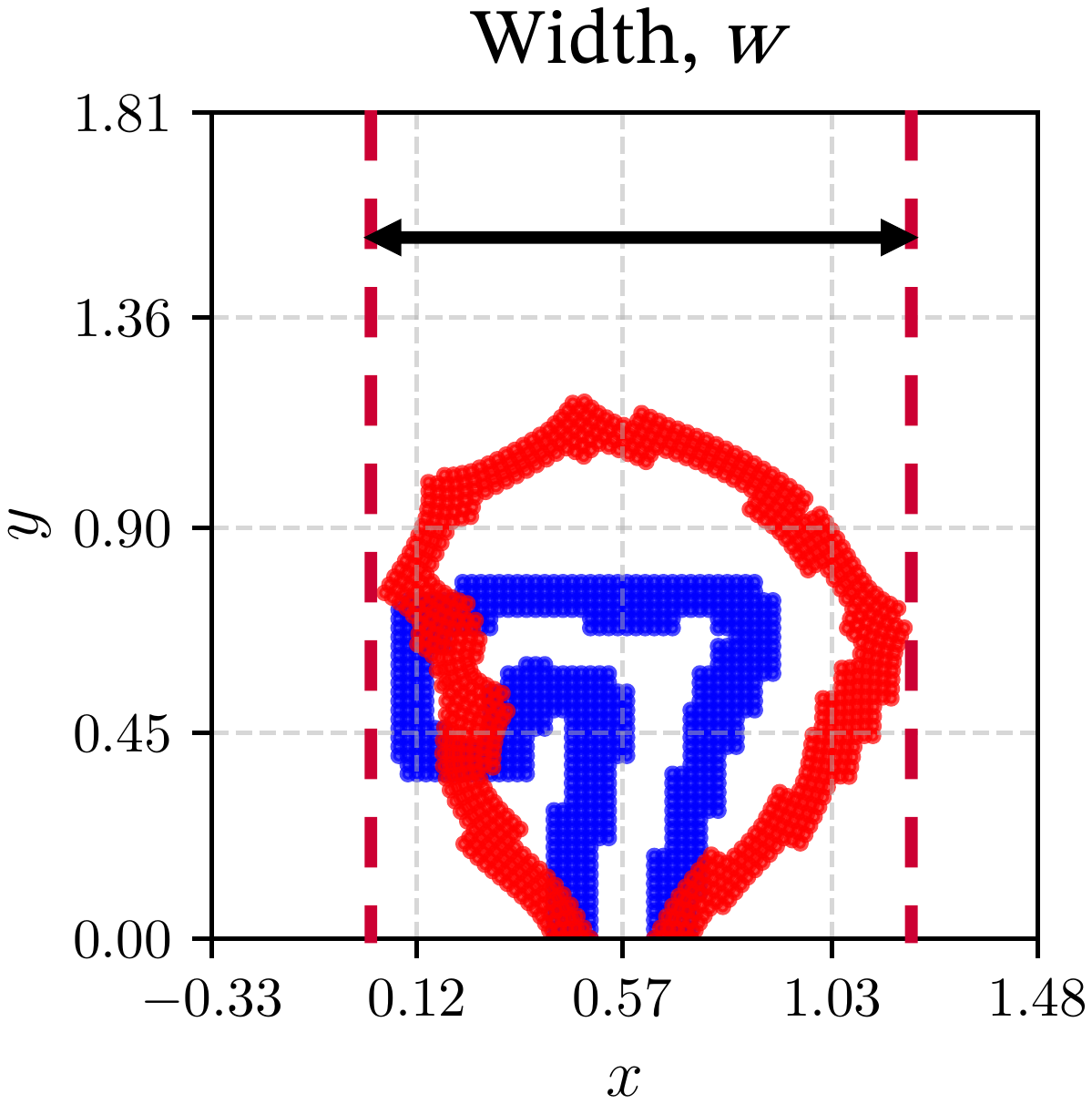}
        \caption{}
        \label{fig:desc_width}
    \end{subfigure}
    \hfill
    \begin{subfigure}[b]{0.19\textwidth}
        \includegraphics[height=3.6cm]{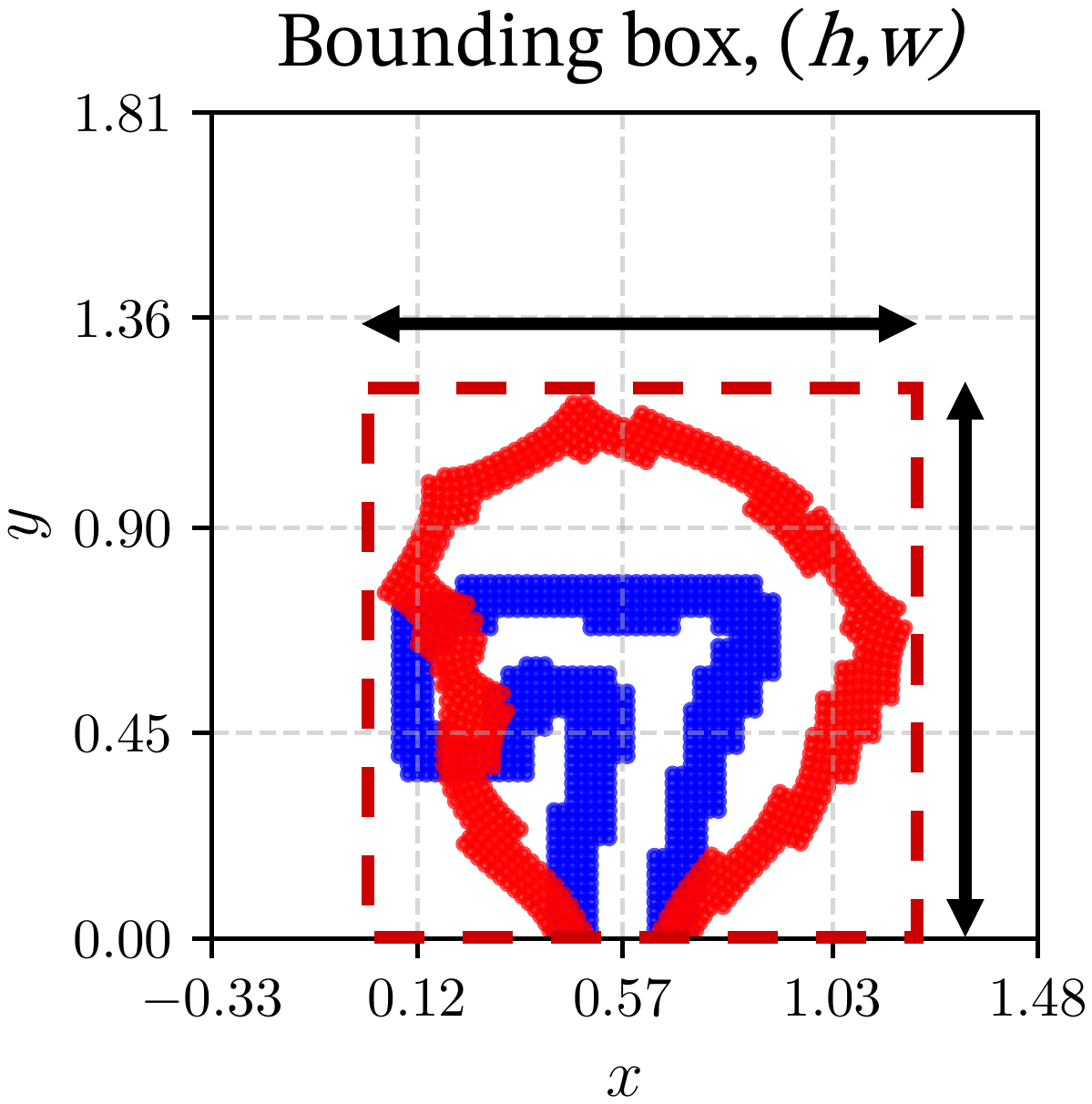}
        \caption{}
        \label{fig:desc_box}
    \end{subfigure}
    \hfill
    \begin{subfigure}[b]{0.19\textwidth}
        \includegraphics[height=3.6cm]{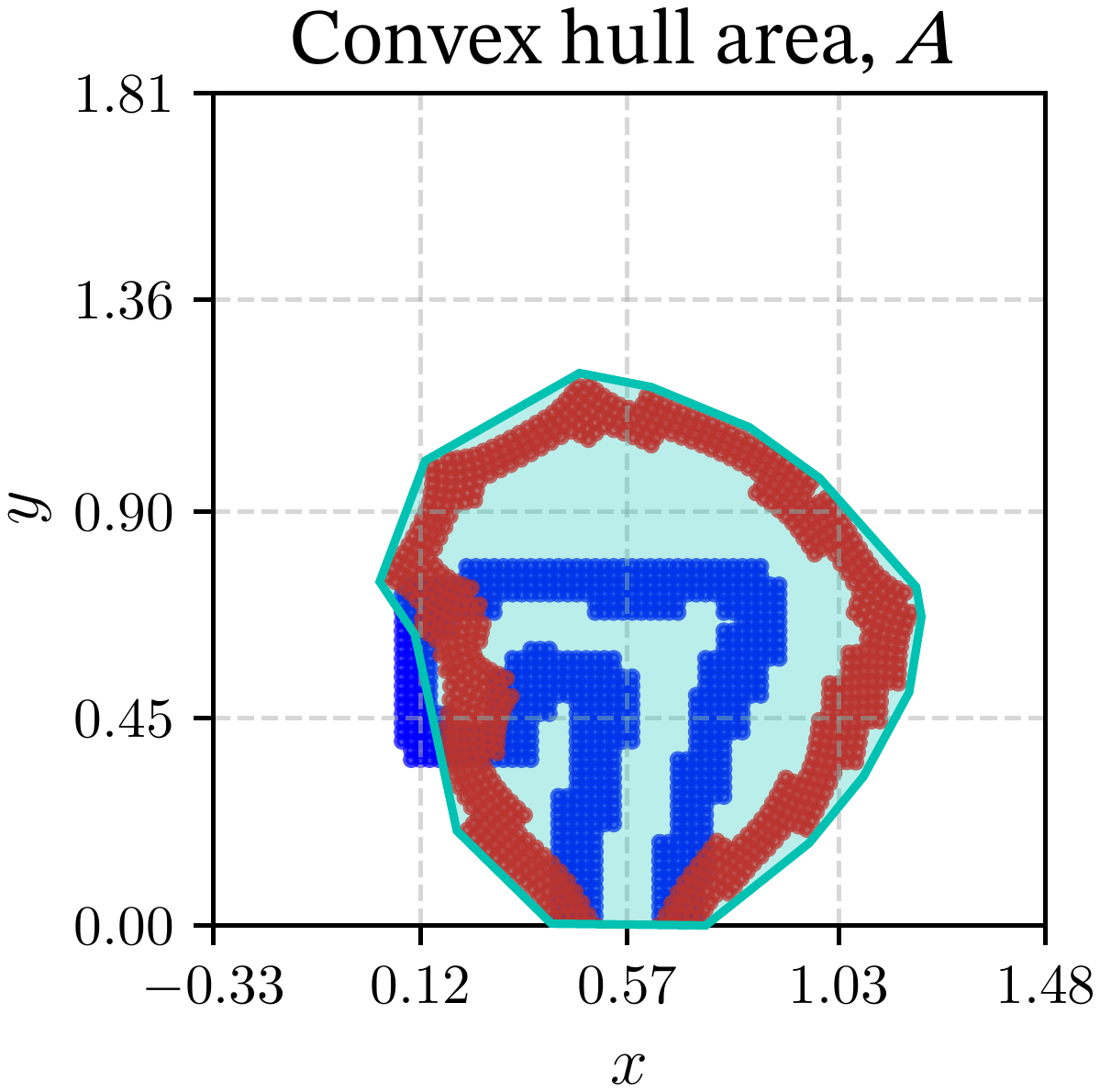}
        \caption{}
        \label{fig:desc_area}
    \end{subfigure}
    \hfill
    \begin{subfigure}[b]{0.19\textwidth}
        \includegraphics[height=3.6cm]{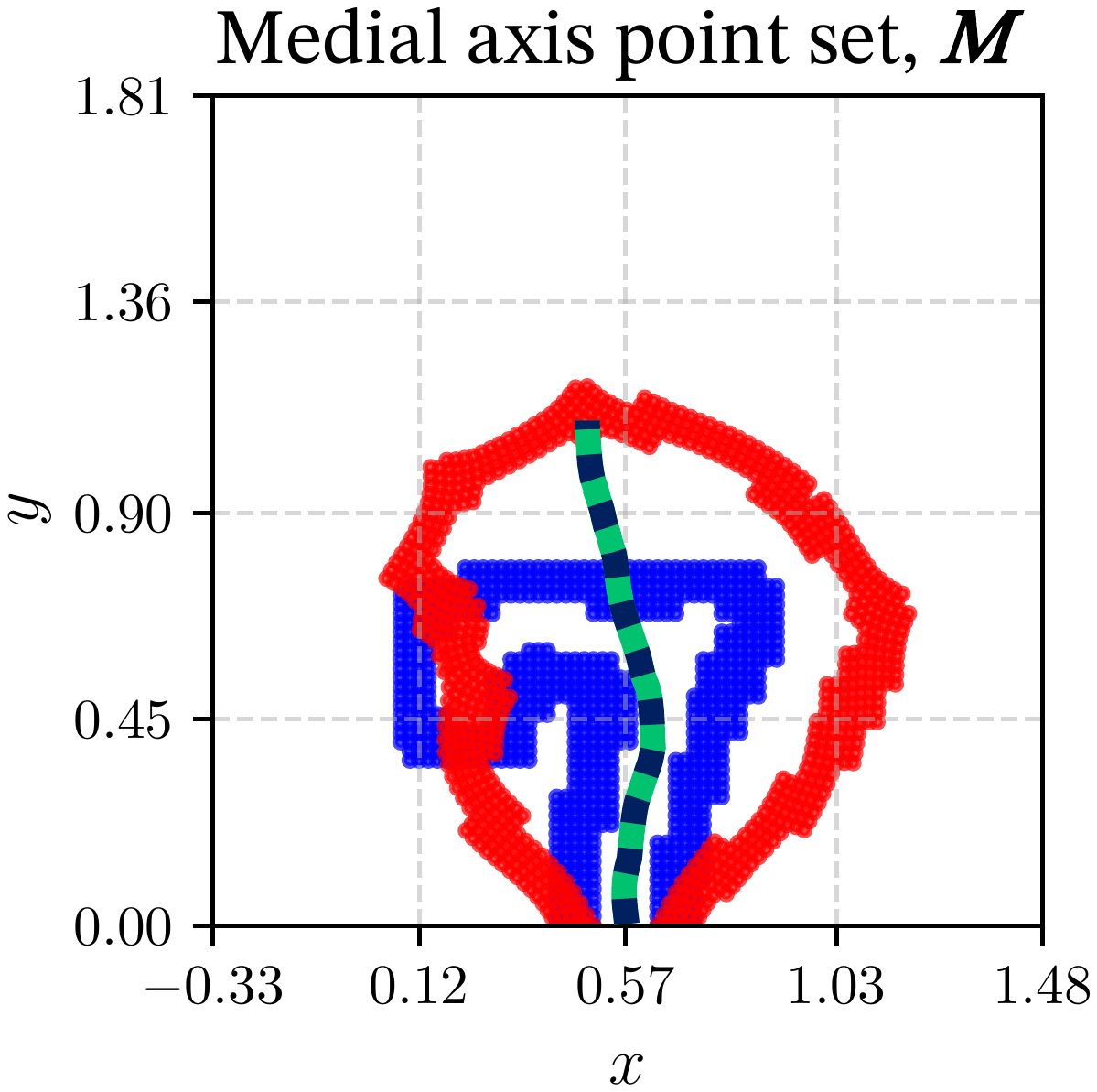}
        \caption{}
        \label{fig:desc_medial}
    \end{subfigure}
    \caption{Calculation of five target descriptors for the inflatable structures: (a) height, (b) width, (c) bounding box, (d) convex hull area, and (e) medial axis.}
    \label{fig:descriptors_row}
\end{figure}

Thus, we extract a set of geometric descriptors from the deformed configurations of each simulated inflatable structure as shown in Fig.~\ref{fig:descriptors_row}. 
These descriptors provide a simplified yet informative representation of the deformed shape and are used as conditioning targets during training and sampling. 
The set includes both size and shape features, each with a simple geometric interpretation:
\begin{itemize}
  \item[(i)] \textbf{Height} $h$: a scalar representing the vertical extent of the deformed structure, measured as the difference between the maximum and minimum $y$-coordinates of any point in the simulation.
  \item[(ii)] \textbf{Width} $w$: a scalar representing the horizontal extent, computed similarly as the span of occupied $x$-coordinates.
  \item[(iii)] \textbf{Bounding box dimensions} $(h, w)$: a 2D tuple combining the height and width, which together encode the overall scale and aspect ratio of the deformed configuration.
  \item[(iv)] \textbf{Convex hull area} $A$: a scalar denoting the area enclosed by the convex hull of the deformed structure's boundary. This captures the global footprint of the shape.
\item[(v)] \textbf{Medial axis point set} $\mathbf{M}$: an ordered set of 2D coordinates representing the approximate centerline of the deformed shape. 
This point set is derived from the convex hull of the inflated structure and captures local geometric features such as curvature that are not reflected in global size measures. 
The medial axis is extracted by identifying midpoints of line segments spanning the convex hull at fixed height increments.

\quad The medial axis in this work is represented by 30 points. 
The fixed increments are calculated by subdividing the height $h$ of the deformed structure by the number of points representing the axis.
The points are introduced in the set in increasing height order. It is noted that this ordered set is not permutation invariant -- the order of inputs is meaningful.
The resulting point set is transformed so that the first (bottom) point is aligned with the axes' origin.
We also check that the resulting point set is contained within a connected, convex subset of the inflated geometry. 
This ensures that the definition of the medial axis remains consistent across samples even in the presence of non-convexities or branching. 
\end{itemize}
These descriptors are automatically computed from the simulated displacement field at the final time step and serve as a target representation for training the generative diffusion model. Although not exhaustive, the chosen descriptors balance expressiveness and computational simplicity, and enable structured generation of designs with targeted deformation characteristics. A single data sample in our dataset is defined as a pair comprising a structural basis (i.e., a grayscale image input) and the corresponding set of descriptors computed from its deformed configuration. Thus, we calculate the five above descriptors for the 6265 simulations performed as described in the previous section.

\begin{rmk}
We deliberately fix the image preprocessing, simulation setup, and descriptor extraction procedure across all database and generated samples to ensure consistency and simplify the learning task for the generative model. 
By using the same approach to label material regions, apply boundary conditions, and simulate internal pressurization, we ensure that each structural input corresponds to a well-defined and physically meaningful deformation. 
This standardization allows the diffusion model to operate entirely in the image domain without requiring explicit representations of meshes or simulation-specific data structures. The model implicitly learns to associate structural inputs -- represented as images -- with deformation behaviors observed in the simulation. 
\end{rmk}

\subsection{Boundary condition-preserving data augmentation}
\label{sec:bc}

To increase the diversity of the dataset without altering the outcome of the associated simulations, we apply data augmentation to the structural basis images so that the boundary conditions are respected. 
Not all standard augmentation procedures used in computer vision, such as arbitrary rotations or rescaling, would be suitable here because they would change the geometry in ways that violate the assumptions of the fixed preprocessing and simulation pipeline. 
For example, a rotated input would generate a different mesh and boundary mapping, breaking the correspondence with the original simulation result and a new simulation would have to be re-run.
Thus, applied transformations are restricted to preserve compatibility with the fixed preprocessing and simulation pipeline. 
In particular, we allow vertical and horizontal content shifts as well as reflections about the y-axis, such that (a) the undeformed configuration size and shape remain practically identical and (b) the displacement boundary conditions remain unchanged; the bottom boundary of the structure remains parallel to the bottom image frame.

Applied vertical and horizontal pixel shifts reposition the structure within the frame while ensuring that the base of the geometry remains aligned with the bottom boundary where displacement constraints are applied. 
Reflections about the vertical axis maintain the topology and loading configuration, producing valid symmetric variants that respect both the material domain labeling and the simulation setup.
The corresponding descriptors are carefully adjusted to match the augmented images, ensuring that the model learns the correct associations between structural inputs and their deformation behaviors.
The height, width, bounding box, and area descriptors remain identical for the transformed configurations for all types of transformations.
The calculated medial axis point set remains the same for all pixel shift transformations and is reflected about the vertical axis for the reflection transformations.

\begin{figure}[htbp]
  \centering

  \begin{subfigure}[t]{0.12\textwidth}
    \includegraphics[width=\linewidth]{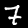}
    \caption*{\footnotesize Original}
  \end{subfigure}\hspace{0.4em}%
  \begin{subfigure}[t]{0.12\textwidth}
    \includegraphics[width=\linewidth]{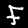}
    \caption*{\footnotesize Flipped}
  \end{subfigure}\hspace{0.4em}%
  \begin{subfigure}[t]{0.12\textwidth}
    \includegraphics[width=\linewidth]{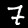}
    \caption*{\footnotesize $3$ pixels right}
  \end{subfigure}\hspace{0.4em}%
  \begin{subfigure}[t]{0.12\textwidth}
    \includegraphics[width=\linewidth]{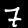}
    \caption*{\footnotesize  $2$ pixels down}
  \end{subfigure}\hspace{0.4em}%
  \caption{Examples of the data-augmentation operations applied to each
           structural basis:  mirror reflection about the vertical axis and pixel
           shifts of up to $\pm3$\,pixels horizontally or vertically.}
  \label{fig:data-augmentation}
\end{figure}

The data augmentation methods are illustrated in Fig.~\ref{fig:data-augmentation}, which shows examples of reflected structures and pixel-shifted variants.
 For each original structural basis,  flip and random vertical/horizontal translations of up to three pixels in each direction were applied in random order, with possible repetition and combination of transformations, while retaining only those shifts that kept the structure entirely within the image frame. 
These operations expanded the training dataset from 6265 to 71616 samples, making the training of the generative framework feasible.

\section{Numerical Experiments}
\label{sec:numerical_experiments}

We evaluate the proposed framework through five independent experiments, each conditioned on a different geometric descriptor of the deformed configuration: height, width, bounding box dimensions, convex hull area, and medial axis coordinates. Separating the experiments by descriptor type test the model’s ability to capture and reproduce different levels of geometric complexity -- from simple scalar targets to high-dimensional, spatially ordered point sets. 
For each descriptor, a separate conditional DDPM is trained and used to generate structural bases for selected target values. The generated structures are then processed through the fixed preprocessing, simulation, and postprocessings, and the resulting deformations are compared against the targets to assess accuracy and variability. This section first outlines the shared model setup and training protocol, followed by descriptor-specific results.

\subsection{Setup and training of the DDPM framework}

In this work, the reverse denoising process is modeled using a U-net architecture tailored to the small spatial resolution and grayscale nature of the input data. 
The design follows a simplified variant of the implementation in \citet{nichol2021improved}, originally developed for $64\times 64$ RGB images from ImageNet \citep{deng2009imagenet} and CIFAR-10 \citep{krizhevsky2009learning}. 
Our inputs are single-channel $28\times 28$ grayscale images representing 2D structural bases, and thus a narrower U-net is adopted. 
The network comprises three resolution scales with channel depths of $(1, 64, 128, 256)$ across the encoder and decoder branches. 
Each down- and up-sampling stage includes convolutional layers with residual connections. Temporal information is encoded using sinusoidal positional embeddings, allowing the model to represent each noise level in the diffusion process continuously in a 256-dimensional vector. The diffusion process is discretized into 1000 steps using a cosine noise schedule. Classifier-free guidance is employed during generation with a dropout rate of 0.1 and guidance strength set to 1. 
The model is trained using the simple noise loss augmented by a variational lower bound term scaled by a factor of 0.001 as described in Eq.~\eqref{eq:hybrid_loss}.

Across experiments, we keep the U-net denoising backbone, diffusion hyperparameters, optimizer, number of training epochs, and loss formulation fixed. The only varying component is the context embedding module -- a two-layer feed-forward network that embeds the conditioning descriptor into a 256-dimensional latent space vector to be added to the sinusoidal time embedding. The network architecture is identical across runs: two hidden layers with 200 units and SiLU activations, followed by a linear output layer. The input dimensionality to the network is denoted by $N$, which depends on the descriptor type: $N = 1$ for scalar descriptors (e.g., height, width, or area), $N = 2$ for descriptor tuples (e.g., height and width), and $N = 2 \times 30$ for point set descriptors such as the selected medial axes with 30 points. This design allows consistent conditioning of the generative model across structurally diverse input types.

The diffusion model is trained on the  dataset described in Section~\ref{sec:database}, consisting of pairs of structural basis images and deformation descriptors. 
Each grayscale image is scaled to the range $[-1, 1]$, and each descriptor vector is normalized to $[0, 1]$ using min--max scaling. 
The model is trained for 150 epochs using batches of 100 samples, with shuffling of the samples at each epoch. 
Optimization is performed using the Adam optimizer \citep{kingma2015adam} with  a learning rate of $10^{-4}$. 
Model checkpoints are saved every 10 epochs. 
All training is performed on the Metal Performance Shaders (MPS) backend, enabling efficient utilization of Apple Silicon hardware for accelerated deep learning.

\begin{figure}[htbp]
  \centering

  \begin{subfigure}[t]{0.095\textwidth}
    \includegraphics[width=\linewidth]{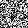}
    \caption*{\small Denoising step 1}
  \end{subfigure}
  \begin{subfigure}[t]{0.095\textwidth}
    \includegraphics[width=\linewidth]{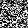}
    \caption*{\small Denoising step 200}
  \end{subfigure}
  \begin{subfigure}[t]{0.095\textwidth}
    \includegraphics[width=\linewidth]{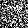}
    \caption*{\small Denoising step 300}
  \end{subfigure}
  \begin{subfigure}[t]{0.095\textwidth}
    \includegraphics[width=\linewidth]{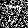}
    \caption*{\small Denoising step 400}
  \end{subfigure}
  \begin{subfigure}[t]{0.095\textwidth}
    \includegraphics[width=\linewidth]{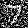}
    \caption*{\small Denoising step 500}
  \end{subfigure}
  \begin{subfigure}[t]{0.095\textwidth}
    \includegraphics[width=\linewidth]{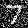}
    \caption*{\small Denoising step 600}
  \end{subfigure}
  \begin{subfigure}[t]{0.095\textwidth}
    \includegraphics[width=\linewidth]{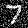}
    \caption*{\small Denoising step 700}
  \end{subfigure}
  \begin{subfigure}[t]{0.095\textwidth}
    \includegraphics[width=\linewidth]{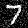}
    \caption*{\small Denoising step 800}
  \end{subfigure}
  \begin{subfigure}[t]{0.095\textwidth}
    \includegraphics[width=\linewidth]{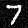}
    \caption*{\small Denoising step 900}
  \end{subfigure}
  \begin{subfigure}[t]{0.095\textwidth}
    \includegraphics[width=\linewidth]{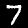}
    \caption*{\small Denoising step 1000}
  \end{subfigure}

  \caption{Denoising process from step 1 to 1000 for a single generated sample. }
  \label{fig:denoising_steps}
\end{figure}
For each numerical experiment, we train a separate DDPM model conditioned on a specific combination of deformation descriptors. 
This results in five distinct models, each corresponding to a different input type.
During sampling, a target descriptor is provided as input, and the model generates structural basis images by running the learned denoising process in reverse, starting from  Gaussian noise. 
Fig.~\ref{fig:denoising_steps} illustrates the evolution of a sample through the generative process, showing representative intermediate outputs from timestep 0 (pure noise) to timestep 1000 (fully denoised structure). 
The generated structural basis is then passed through the same preprocessing, simulation, and postprocessing pipeline as described in Section~\ref{sec:database}. The results of this procedure are presented and analyzed in the following sections.
\subsection{Generating structures with target scalar descriptors}

\begin{figure}[H]
    \centering
    \begin{subfigure}[b]{0.23\textwidth}
        \includegraphics[width=\linewidth]{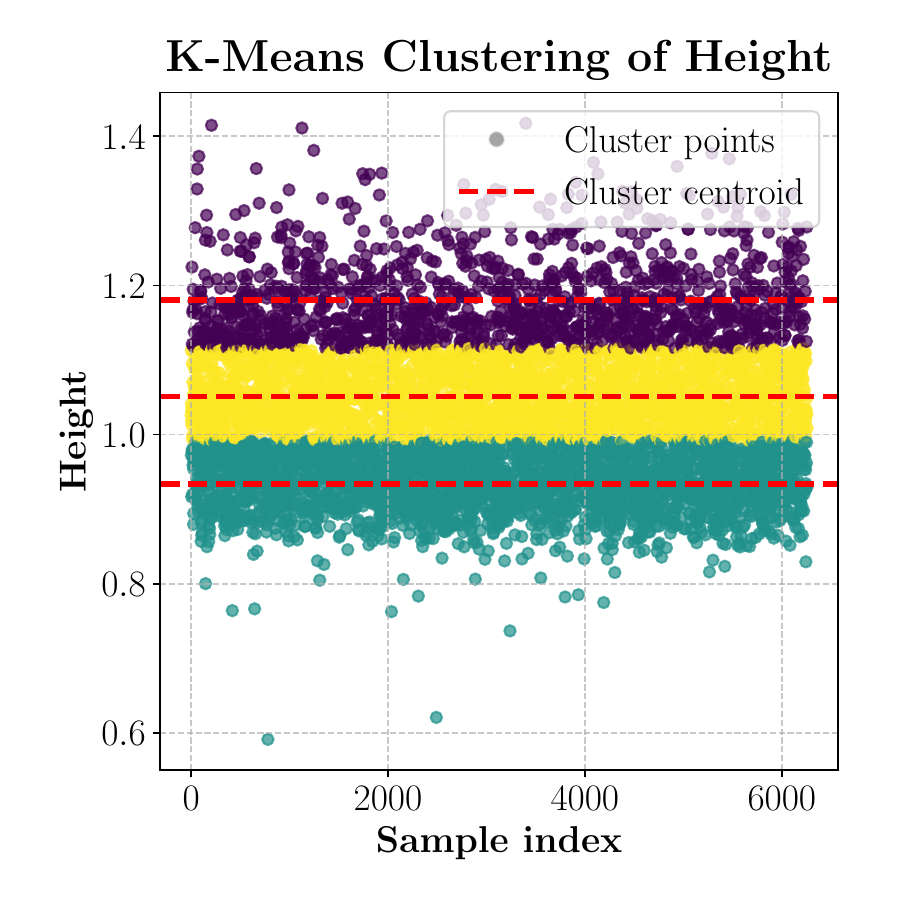}
        \caption{}
        \label{fig:kmeans_height}
    \end{subfigure}
    \hfill
    \begin{subfigure}[b]{0.23\textwidth}
        \includegraphics[width=\linewidth]{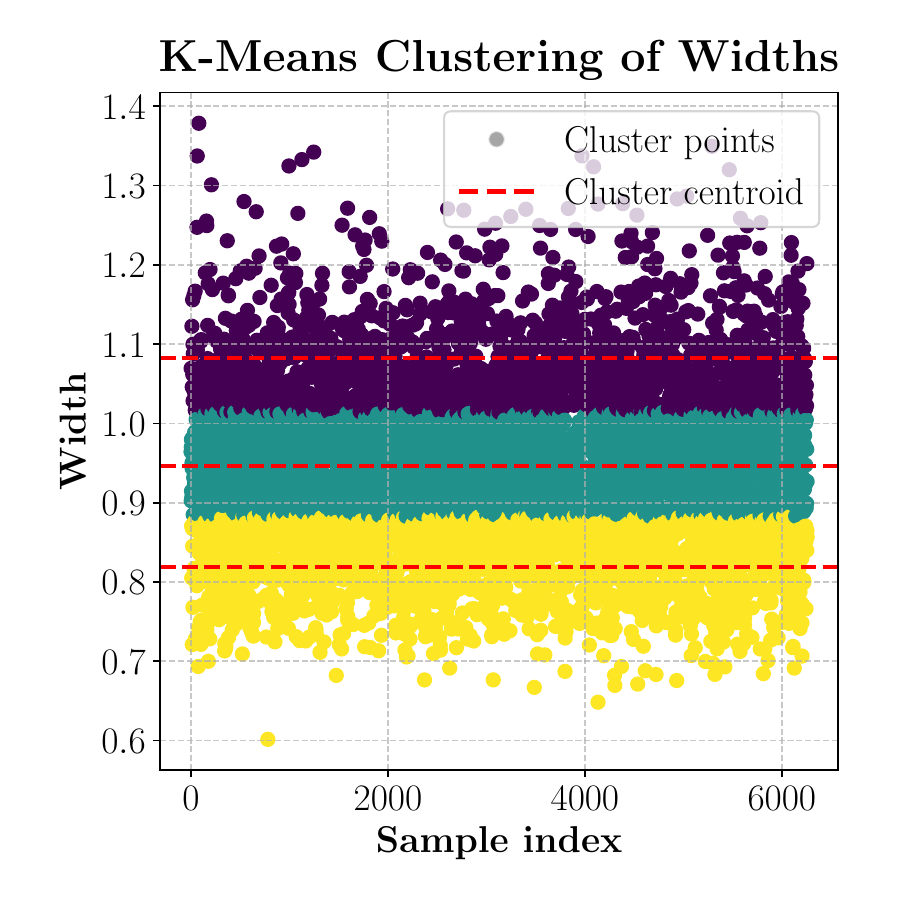}
        \caption{}
        \label{fig:kmeans_width}
    \end{subfigure}
    \hfill
    \begin{subfigure}[b]{0.23\textwidth}
        \includegraphics[width=\linewidth]{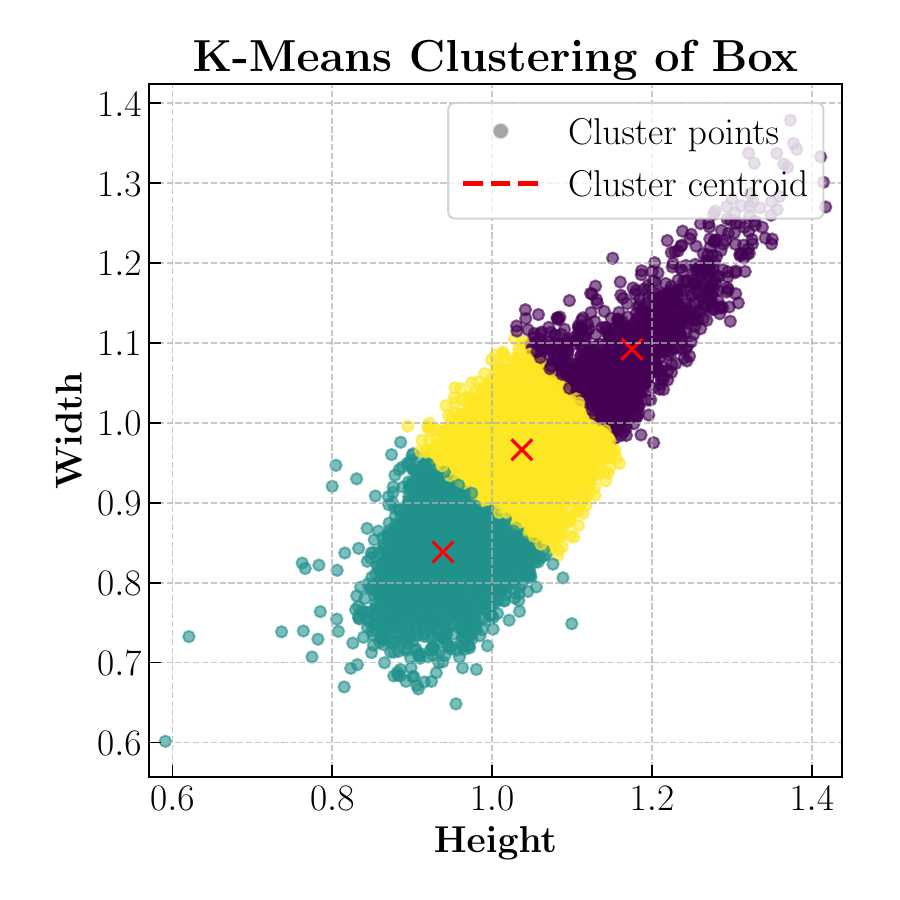}
        \caption{}
        \label{fig:kmeans_box}
    \end{subfigure}
    \hfill
    \begin{subfigure}[b]{0.23\textwidth}
        \includegraphics[width=\linewidth]{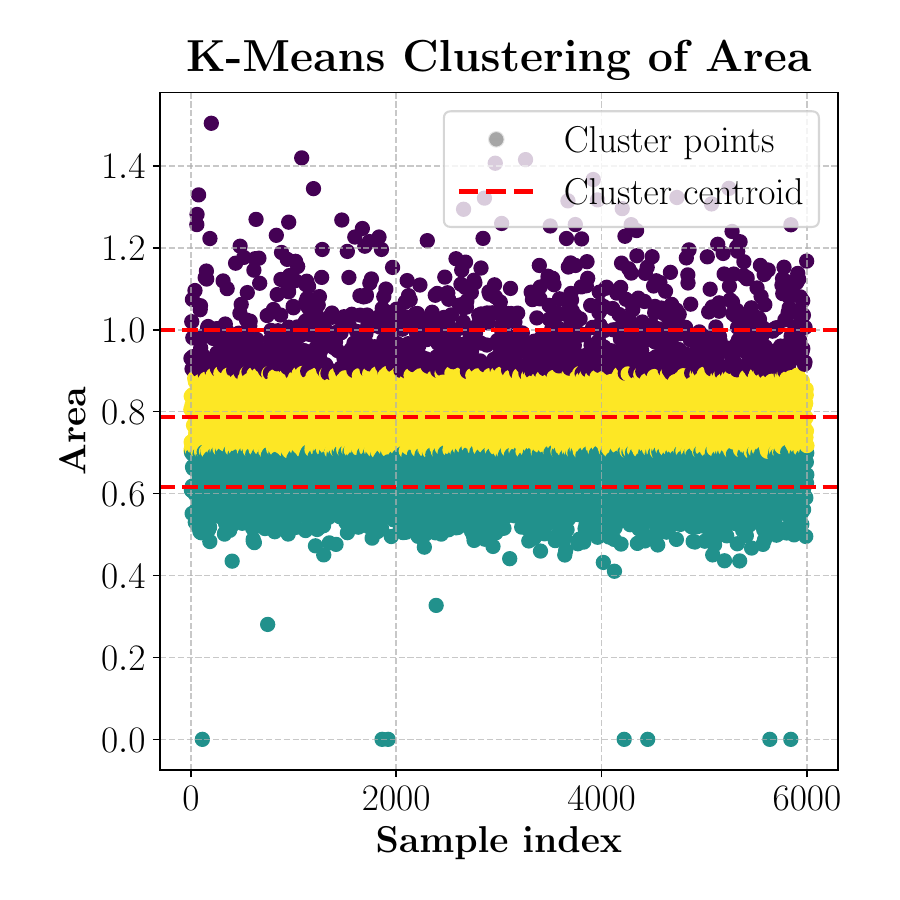}
        \caption{}
        \label{fig:kmeans_area}
    \end{subfigure}
    \caption{K-means clustering results for (a) height, (b) width, (c) bounding box, and (d) area descriptors. Three representative cluster centroids are selected per descriptor for unbiased validation of the algorithm in the numerical experiments.}
    \label{fig:kmeans_row}
\end{figure}

In the first four numerical experiments, we target scalar geometric descriptors of the deformed structure: height, width, aspect ratio, and convex hull area. For each descriptor, we train a separate DDPM model as outlined in the previous sections. To ensure representative and unbiased sampling across the dataset, we apply $k$-means clustering to the distribution of each target descriptor and select three clusters per descriptor. For scalar descriptors such as height, width, aspect ratio, and convex hull area, the clusters correspond to representative low, medium, and high value ranges. For the bounding box descriptor, which is a tuple of height and width, the clustering yields three representative combinations capturing the joint variability of the two quantities. In all cases, we use the centroids of the resulting clusters as target inputs for conditional generation. These clusters and their centroids are visualized in Fig.~\ref{fig:kmeans_row} and the corresponding values are summarized in Table~\ref{tab:descriptor_targets}.

\begin{figure}[H]
    \centering
    \includegraphics[width=0.98\linewidth]{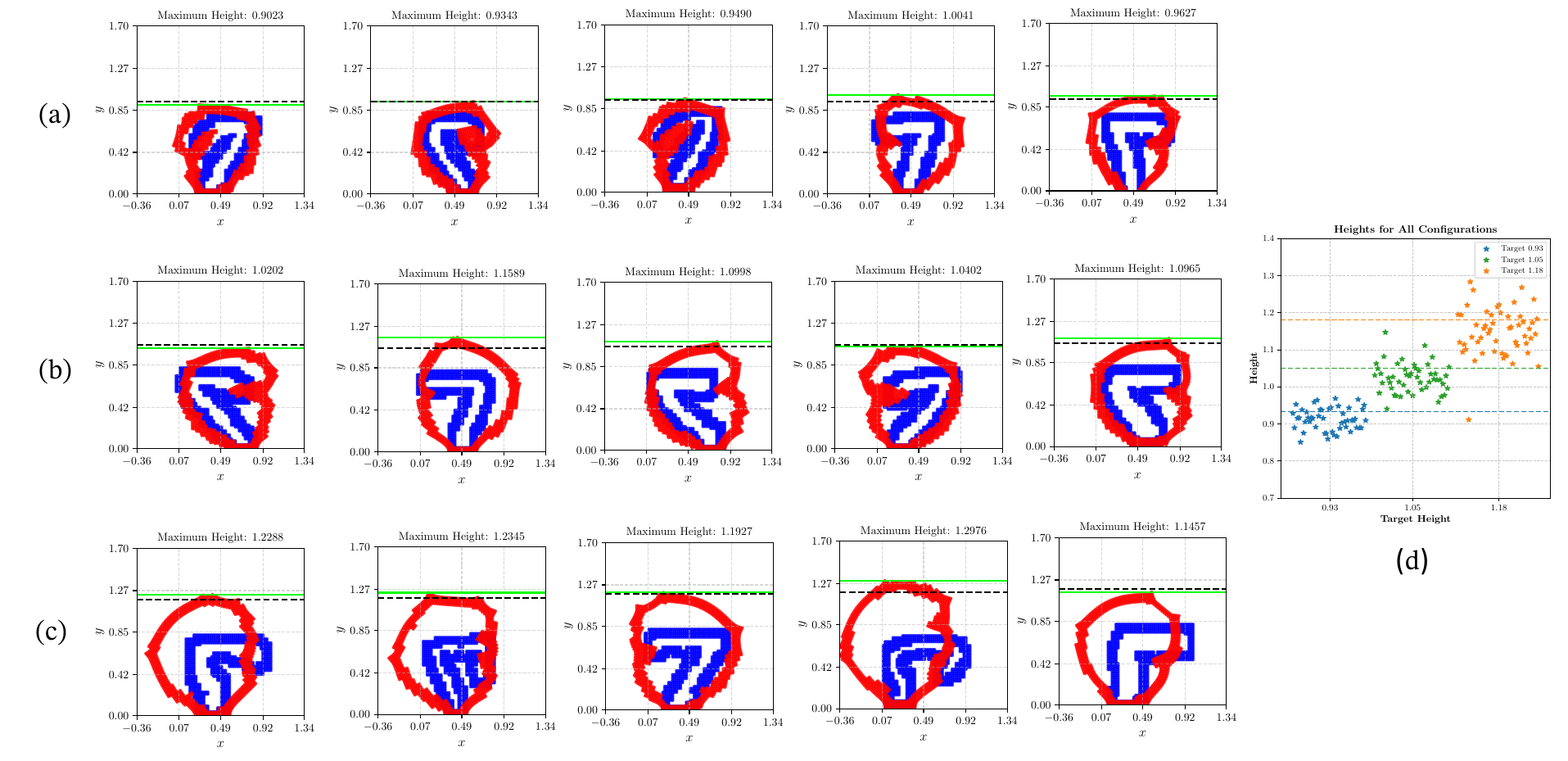}
    \caption{Five generated samples with a target height of (a) 0.93, (b) 1.05, and (c) 1.18. (d) Calculated height for 50 generated samples per target height descriptors.}
    \label{fig:height}
\end{figure}

\begin{figure}[H]
    \centering
    \includegraphics[width=0.98\linewidth]{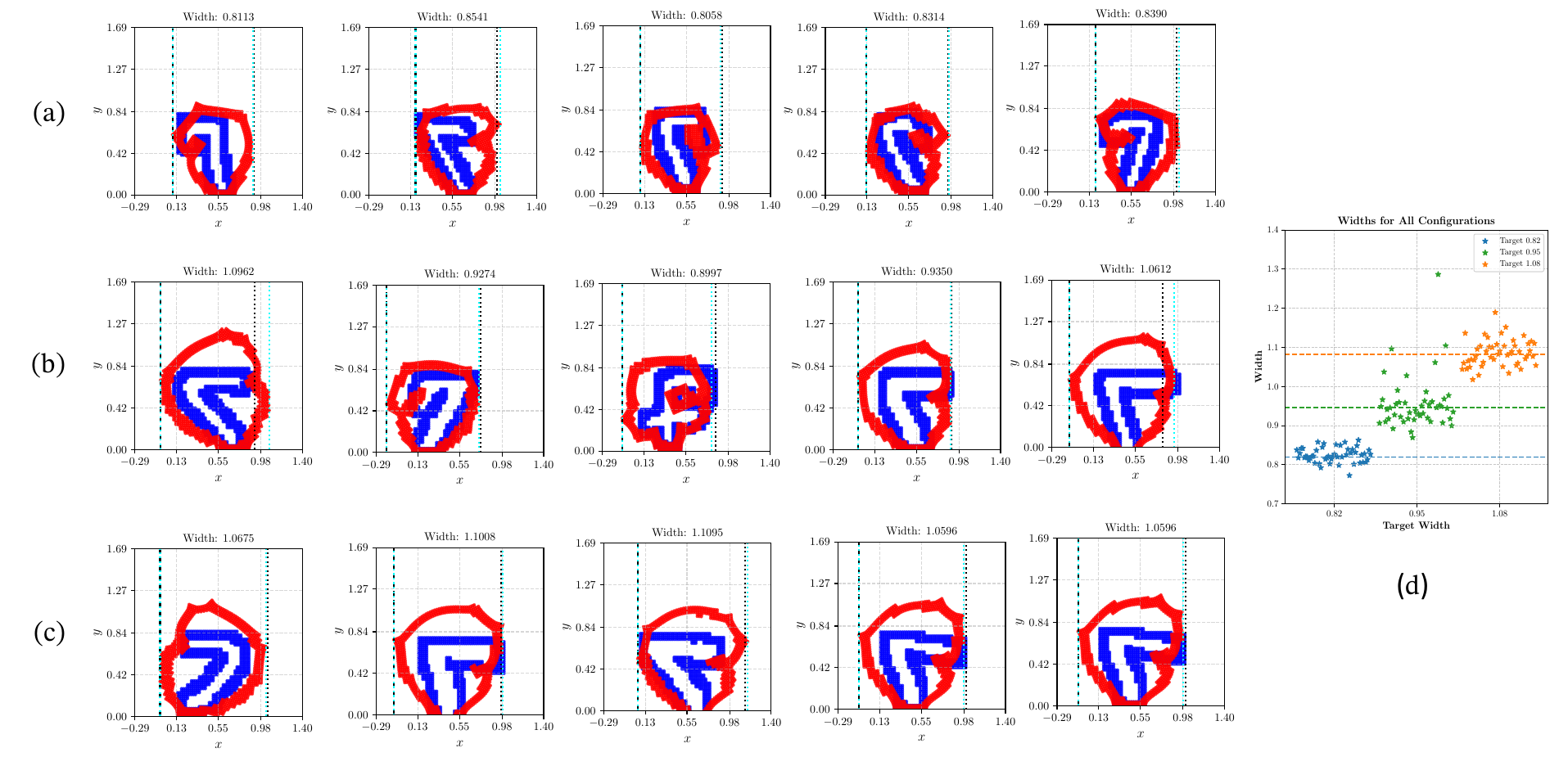}
    \caption{Five generated samples with a target width of (a) 0.82, (b) 0.95, and (c) 1.082. (d) Calculated width for 50 generated samples per target width descriptors.}
    \label{fig:width}
\end{figure}

\begin{figure}[H]
    \centering
    \includegraphics[width=0.98\linewidth]{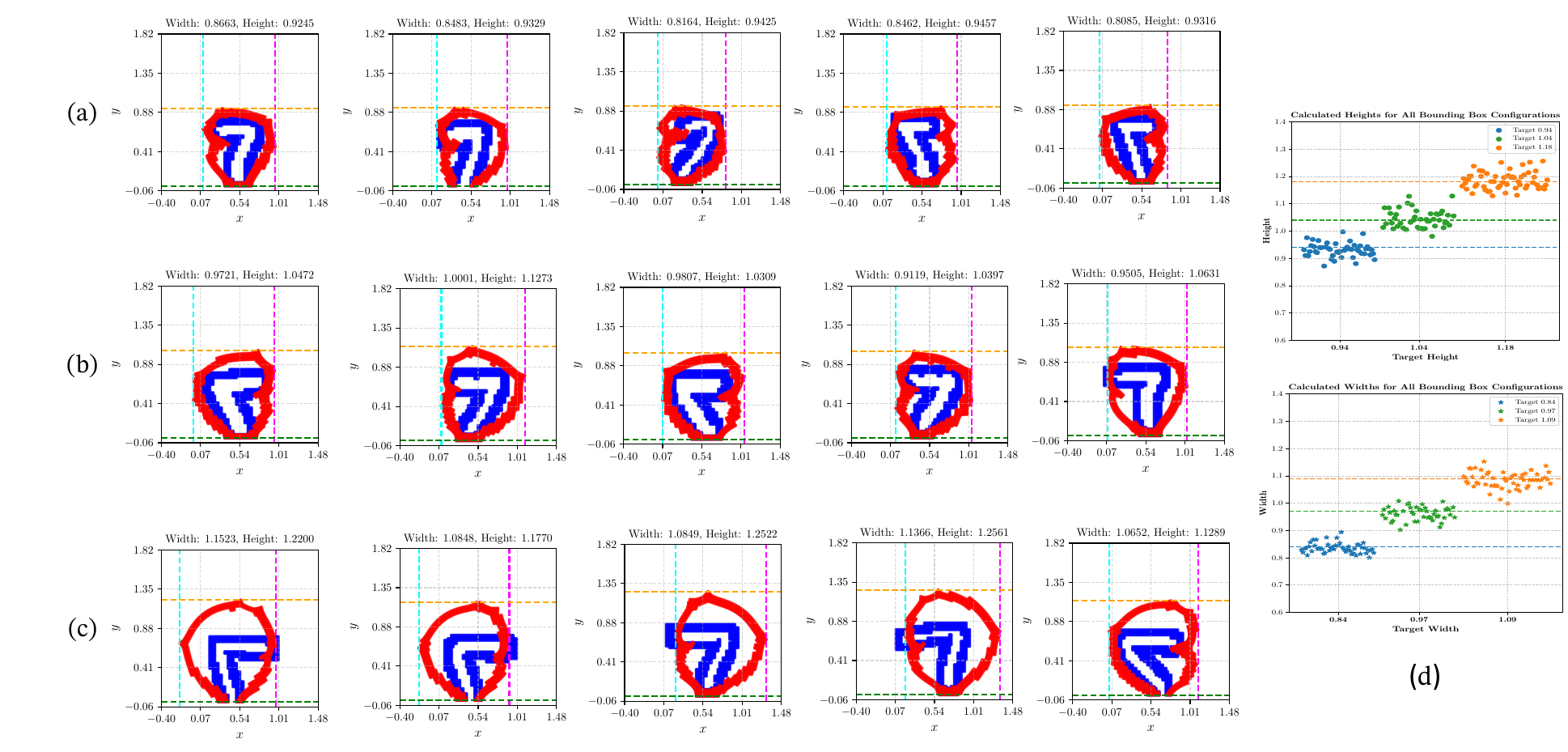}
    \caption{Five generated samples with a target bounding box of (a) Height = 1.18, Width = 1.09, (b) Height = 1.04, Width = 0.97, and (c) Height = 0.94, Width = 0.84. (d) Calculated bounding box for 50 generated samples per target bounding box descriptors.}
    \label{fig:box}
\end{figure}

\begin{figure}[H]
    \centering
    \includegraphics[width=0.98\linewidth]{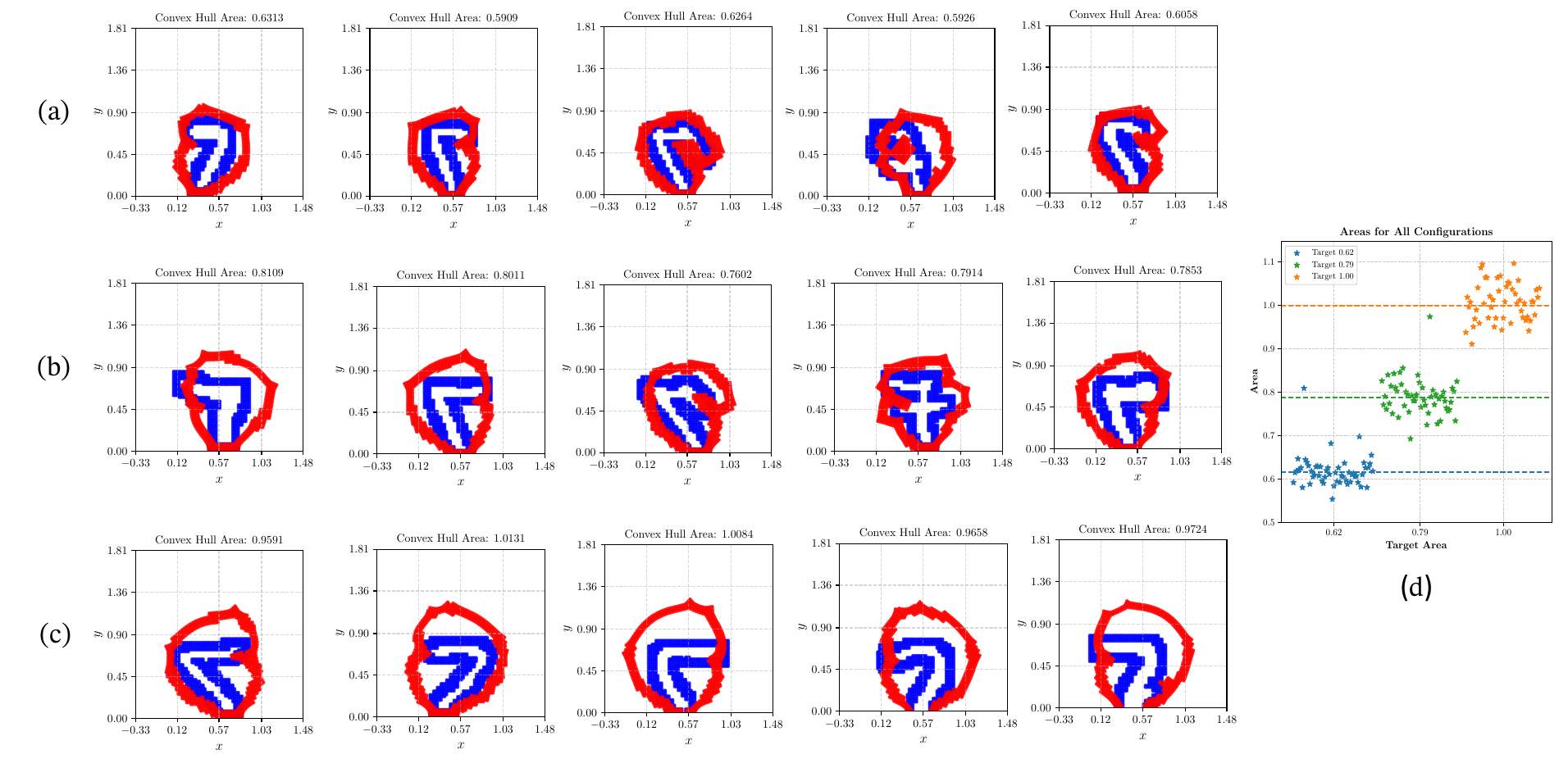}
    \caption{Five generated samples with a target area of (a) 0.62, (b) 0.8, and (c) 0.9. (d) Calculated area for 50 generated samples per target area descriptors.}
    \label{fig:area}
\end{figure}

\begin{table}[h]
\centering
\caption{Selected target values (cluster centroids) for each scalar descriptor used in the numerical experiments.}
\begin{tabular}{ll}
\toprule
\textbf{Descriptor} & \textbf{Target Values} \\
\midrule
Height ($h$) & 0.93,\quad 1.05,\quad 1.18 \\
Width ($w$) & 0.82,\quad 0.95,\quad 1.08 \\
Bounding box ($h$, $w$) & 
\begin{tabular}[t]{@{}l@{}}
(0.94, 0.84),\quad $h/w = 1.12$ \\
(1.04, 0.97),\quad $h/w = 1.07$ \\
(1.18, 1.09),\quad $h/w = 1.08$
\end{tabular} \\
Convex hull area ($A$) & 0.62,\quad 0.79,\quad 1.00 \\
\bottomrule
\end{tabular}
\label{tab:descriptor_targets}
\end{table}
For each selected target value/cluster centroid, we generate 50 structural basis samples conditioned on the descriptor. These samples are passed through the fixed preprocessing, simulation, and postprocessing pipeline detailed in Section~\ref{sec:database}, including image processing, FEA simulation, and descriptor evaluation. Figures~\ref{fig:height},~\ref{fig:width},~\ref{fig:box}, and \ref{fig:area}  summarize the results for each scalar descriptor. On the left side of each figure, we display five representative examples of undeformed and deformed configurations corresponding to the selected targets. 
For height, width, and bounding box descriptors, visual indicators (e.g., vertical/horizontal lines or bounding rectangles) are overlaid to illustrate agreement between the generated and target values. For convex hull area, the calculated area values are annotated directly on each image instead of using graphical overlays. On the right side of each figure, we present the distribution of all 50 generated samples for each target, showing how the achieved descriptor values cluster around the centroid.

While some variability is expected due to the stochastic nature of the sampling process and the absence of explicit descriptor loss terms in the training objective, the generated results consistently concentrate around the desired target values. 
This outcome confirms that the diffusion model effectively learns the statistical relationship between descriptor values and structural basis configurations, even without direct supervision on descriptor accuracy.
 However, in settings where precise control over the generated descriptor values is critical additional mechanisms may be introduced to refine the output. 
 These include surrogate models trained to predict descriptor values from generated structures, enabling rapid post-generation ranking and filtering, or direct simulation-based selection, where candidate samples are evaluated using the full physics-based pipeline and selected based on their agreement with the target. 
 Such methods were explored in work by \citet{vlassis2023denoising}, but were not deemed necessary here, as the goal of this study is to establish that conditional generation around meaningful target ranges is achievable using scalar descriptors, and they will be considered in future work.


\subsection{Generating structures with target medial axes}

To evaluate the capacity of the generative framework to handle more complex geometric conditioning beyond scalar or tuple-based descriptors, we conduct a fifth numerical experiment targeting medial axis representations. 
Unlike the previous experiments, where scalar values or low-dimensional tuples are used to define conditioning targets, medial axes represent higher-dimensional, spatially varying geometric objects that characterize the internal structure of a shape and capture local curvature and branching information that cannot be represented by global size measures alone.

Since medial axes cannot be easily clustered using standard $k$-means techniques without compromising the physical meaning of the axis, we instead manually select three qualitatively distinct medial axis inputs that span a diverse subset of the design space in terms of both size and shape. 
Each selected medial axis serves as a conditional input to generate 50 structural basis samples. These samples are processed through the same pipeline described in Section~\ref{sec:database}, including preprocessing, FEA simulation, and postprocessing. 
The medial axes of the deformed structures are extracted and compared against the target axes to assess the model's ability to generate geometries that align with complex shape descriptors.

\begin{figure}[H]
    \centering
    \includegraphics[width=0.98\linewidth]{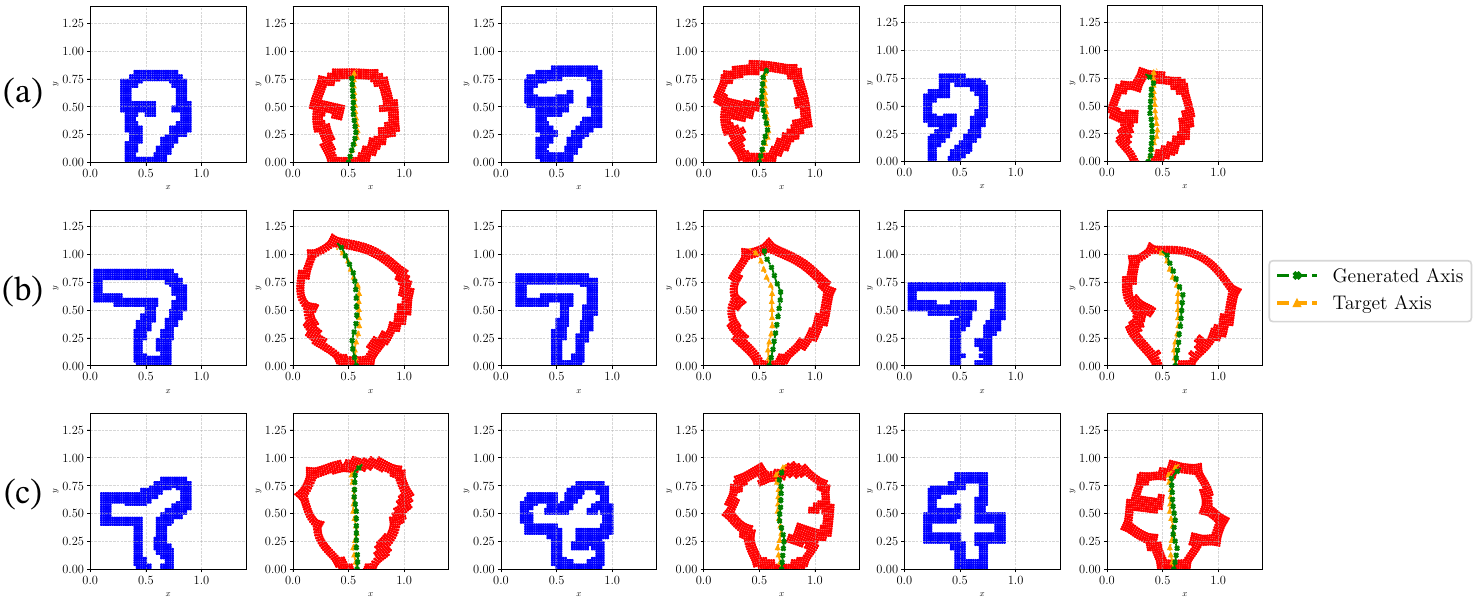}
    \caption{Three generated samples for three different sets of target medial axis coordinates.}
    \label{fig:medial}
\end{figure}

\begin{figure}[H]
    \centering
    \includegraphics[width=0.98\linewidth]{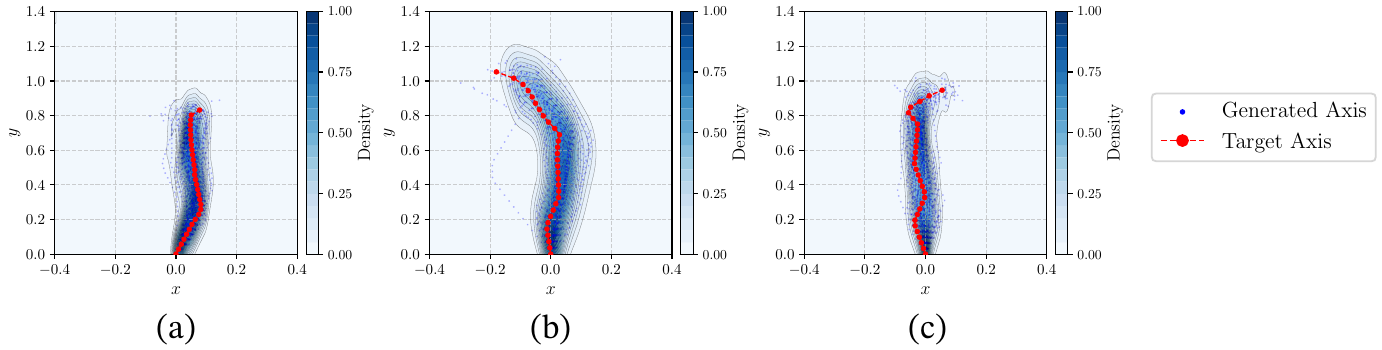}
    \caption{Normalized density distribution of generation results for three target medial axes. 50 configurations were generated per target.}
    \label{fig:medial_axis_all}
\end{figure}

Figures~\ref{fig:medial} and~\ref{fig:medial_axis_all} summarize the results of the medial axis experiments. In Fig.~\ref{fig:medial}, we show three representative generated samples for each of the three selected target medial axes. For each case, we present the undeformed generated configuration, its corresponding deformed configuration, the predicted medial axis extracted from the deformed structure, and the reference target medial axis for visual comparison. This allows qualitative evaluation of how closely the generated geometries adhere to the target structural layout. In Fig.~\ref{fig:medial_axis_all}, we present normalized density estimates of the medial axes extracted from fifty generated configurations conditioned on each target. Each subplot illustrates the distribution and variability of the predicted medial axes relative to the original target axis, with darker regions indicating higher generation probability and highlighting the axes the model is most likely to produce. Despite the high-dimensional and spatially structured nature of the medial axis representation, the density maps demonstrate consistent alignment with the target inputs and limited spread in high-density regions, indicating that the generative model captures and reproduces the dominant geometric trends in the target shapes.

This numerical experiment establishes that the generative framework can be conditioned on more complex high-dimensional inputs, such as point-set representations of medial axes. This confirms the feasibility of controlling the generation process using modalities beyond scalar or low-dimensional tuple descriptors. The results suggest that the framework can be extended to accommodate more complex conditioning inputs -- including image-based or graph-based shape representations -- provided they can be embedded into the context vector, with the medial axis case serving as a proof-of-concept for conditioning on structured, ordered spatial data. In future work, we will also explore conditioning on physical response fields, such as strain or stress distributions, to enable inverse design tasks that incorporate both geometric and mechanical performance criteria and potentially combine multiple high-dimensional descriptors to achieve multi-objective design targets.

\FloatBarrier

\section{Conclusion} 
\label{sec:conclusion}

We presented a generative framework for the inverse design of inflatable structures with programmable deformations using denoising diffusion probabilistic models. 
This work addresses the challenge of efficiently generating physically valid, manufacturable designs for complex deformation targets, which remains a bottleneck in the design of inflatable and highly deformable structures.
The method conditions generation on geometric descriptors of the target deformed configuration, including scalar measures, bounding box tuples, and medial axis representations, enabling controllable synthesis of undeformed structures that deform predictably under fixed boundary conditions. A fixed preprocessing, simulation, and post-processing pipeline ensures compatibility between generated samples and high-fidelity finite element analysis, allowing training on simple grayscale inputs while targeting complex mechanical responses. 
The proposed formulation demonstrates how descriptor-based conditioning in a diffusion framework can bridge the gap between abstract geometric targets and simulation-compatible designs. 
This framework is designed to be adaptable to multiple design domains as it operates on simple structural representations (structural bases) that can be easily defined or extended for different applications.

Results across descriptor types demonstrate the framework’s capacity to capture statistical relationships between geometric targets and structural configurations, supporting both low- and high-dimensional conditioning. The present work focuses on pressure-driven actuation and a limited set of descriptors, and is restricted to a single-material neo-Hookean constitutive model and a single family of baseline geometries. However, heterogeneous material distributions, other physics such as plasticity, and different undeformed configurations can be considered with the appropriate representation of a structural basis. 
For example, spatially varying elasticity coefficients could be encoded as simple as using grayscale or RGB images, directly extending the current framework. 
The approach can be extended to conditioning on richer geometric descriptors, such as shape contours or images, or mechanical fields, such as strain or stress distributions, and to alternative actuation methods. Future work will also investigate integrating surrogate models to control the generation of structures with targeted behaviors or direct descriptor loss terms to improve precision in matching target values, such as through multi-objective conditioning to capture multiple performance metrics simultaneously.

\bibliographystyle{plainnat}
\bibliography{main}

\end{document}